\documentclass[]{revtex4}

\pdfoutput=1

\usepackage{slashed}
\usepackage{mathtools}
\usepackage{hyperref}
\usepackage{amsmath}
\usepackage{amssymb}
\usepackage{latexsym}
\usepackage{amsfonts}
\usepackage{graphicx}
\usepackage{setspace}
\usepackage{braket}

\def\hhref#1{\href{http://arxiv.org/abs/#1}{arXiv:#1}} 

\usepackage{color}

\newcommand{\bea}{\begin{eqnarray}}
\newcommand{\ea}{\end{eqnarray}}
\newcommand{\eea}{\end{eqnarray}}

\begin{document}

\linespread{1}

\title{Hawking Radiation via Complex Geodesics}

\author{Cesim K. Dumlu}

\affiliation{Department of Physics, Middle East Technical University, 06800, Ankara, Turkey}
\affiliation{\,}
\affiliation{Extreme Light Infrastructure-Nuclear Physics (ELI-NP), 077125, M\u{a}gurele, Romania}

\email{cesim.dumlu@eli-np.ro}

\begin{abstract}
We describe in detail the quantum tunneling of massive particles from Kerr black hole by using complex trajectories, which are solutions to the Hamilton's equations of motion with imaginary proper time. The trajectories are  smooth and cover the inner and outer horizon regions. Following the worldline approach, we compute the energy flux at the event horizon  as a summation over these complex trajectories. Density of states is given with the aid of Carter's constant and it is shown to be linear in momenta in the leading order, as long as the phase portrait of the system stays uniform. Under this assumption, we obtain the thermal spectrum $\sim (T^{+}_H)^4$. 

\end{abstract}


\maketitle

\section{Introduction}

Since Hawking's original paper, emission of particles from the black hole event horizon  has been a striking prediction of quantum mechanics \cite{hawk}. Tunneling interpretation of Hawking radiation was made transparent by Parikh's and Wilczek's approach, where the imaginary part of the action was calculated by using radial null geodesics\cite{parwil}. Hawking's original result for the tunneling exponent have also been obtained by using the solutions of Hamilton Jacobi equation, and by the complex path approach in various black hole backgrounds\cite{cp, hj}. A comprehensive review of the emission mechanism and its relation to black hole thermodynamics can be found in the works \cite{page1,primer}. The purpose of this paper is to reproduce the emission spectrum of massive particles from the Kerr black hole by using the worldline approach to quantum tunneling. The worldline formalism has been extensively discussed in the literature and has been successfully applied to the analogous vacuum decay phenomenon, the Schwinger effect \cite{wl1, wl7}. In the semiclassical worldline approach the imaginary part of the action is given by the closed, classical trajectory configurations of the system. These trajectories usually appear as the solutions to the Euclidean classical equations of motion, where time is imaginary: $ x^{0} \rightarrow x_\text{E}^{0} = i\,x ^{0}$.  But for relativistic systems, where orbits are parametrized by a proper time parameter, a more natural prescription is to seek for the tunneling orbits with imaginary proper time: $u \rightarrow i u =s $, as proposed by Rubakov \textit{et al} \cite{rubakov}. Thus we look for the closed tunneling orbits in the Kerr geometry via complexified Hamilton's equations of motion in the form,
\bea
i \frac{d p_{\mu}}{d s} = -\frac{\partial\mathcal{H}}{\partial x^{\mu}}, \quad \, i\frac{d x^{\mu}}{d s} = \frac{\partial\mathcal{H}}{\partial p_{\mu}} \nonumber
\eea
where the Hamiltonian is: $\mathcal{H}=1/2\, g^{\mu\nu} p_{\mu}p_{\nu}$. The factor of ``$i$'' above  makes the closed worldlines essentially complex. An important aspect of  complex worldlines is that they smoothly pass from outer to  inner horizon region. Worldlines  are quasiperiodic off the equator or when $p_{\theta}$ is nonvanishing. The ratio of the periods $\tilde{T}/T_1$, where $\tilde{T}$ is the closure period and $T_1$ is the first period, gives the winding number. This number generically shows the number of times that a given trajectory encircles the horizon. As we will discuss in next section, the classical action, $\oint p_r \dot r \, ds $, weighted by the factor $T_1/\tilde{T}$ reproduces  the Hawking temperature, $T^+_H$, which is uniform over the horizon area.   

In processes  where vacuum breaks down under the influence of an external field, the main technical challenge is to compute the imaginary part of the vacuum to vacuum transition amplitude.  Particularly appealing aspect of the worldline approach is that vacuum to vacuum amplitude is naturally related to the closed orbits because it involves a trace. In fact, total decay rate corresponds to a summation over all possible closed trajectory configurations of the system, not just to a single trajectory.  An important consequence of this fact is that path summation over a dense set of tunneling orbits  naturally allows us to  define  the  emittance of black hole at the horizon. In calculating  the energy flux emitted at the horizon, the Carter's constant, $\mathcal{C}$ plays a special role: by using the definition of $\mathcal{C}$, the integral over $p_{\theta}$ and $\theta$ can be integrated to give the \textit{density of states} for fixed energy, $\omega$ and angular momentum, $j$. 	Based on the assumptions that we explain in section III, energy flux can be obtained in the closed form in  the limit of vanishing $\mathcal{C}$ and  it is shown to be proportional to $ (T^{+}_{H})^4$ in the leading order.

The plan of the paper is as follows. In Section II we briefly recall the worldline formalism and perform the semiclassical expansion of the path integral in the phase space, around the complex geodesics. We explain how  the complex geodesics appear and construct the integration cycles that give the tunneling exponent. In Section III we analyze the emittance of black hole as a summation over complex geodesics and calculate the power output at the event horizon in the limit of vanishing Carter's constant and the final Section contains our conclusions.

\section{Worldline Formalism}

Following Schwinger's  prescription\cite{schw}, we consider the vacuum decay amplitude given by the imaginary part of the effective action in the background metric $ (-,+,+,+)$:
\begin{eqnarray} 
P=1-e^{-2 \, \text{Im}\, \Gamma_{\text{eff}}/\hbar}\approx \frac{2}{\hbar} \text{Im}\, \Gamma_{\text{eff}}
\end{eqnarray}
The effective action for Klein-Gordon field is defined as, (henceforth we set $\hbar=G=c=1$)
\begin{eqnarray}
\Gamma_{\rm eff}^{\rm scalar}&=&i\, \ln\, \det\left[D_\mu^2-m^2 \right]=i \,{\rm tr}\, \ln\left[D_\mu^2-m^2\right],
\label{effective}
\end{eqnarray}
where $D_\mu^2=g^{\mu\nu}\,\nabla_{\mu}\nabla_{\nu}$ and tr denotes the trace. The integral representation of $\Gamma_{\rm eff}^{\rm scalar}$ is given with the aid of Schwinger parameter, $T$
\begin{eqnarray}
\Gamma_{\rm eff}^{\rm scalar}&=&-i\int_0^\infty \frac{dT}{T} \,{\rm tr}\,e^{-i\, \frac{1}{2} \left( D_{\mu}^2- m^2 \right)\, T}
\label{eff1}
\end{eqnarray}
A similar definition for the spinor effective action can be given via tetrad formalism\cite{hagels}. Since the only difference between the scalar and spinor spectrum of the emitted particles is the spin statistics of the tunneling amplitude, we concentrate here on the scalar case. Effective action has the usual path integral interpretation when the trace is performed over the closed trajectory configurations,
\begin{equation}
\Gamma^{\rm scalar}_{\rm eff}=-i\int_0^{\infty}\frac{d T}{T} e^{i \frac{1}{2}m^2\, T} \int \sqrt{-g} \,d^4x(0) \int\limits_{x(0)=x(T)}  \mathcal{D}^{4}x\,e^{-iS[x]},
\label{scalar}
\end{equation}
where $S$ is the classical action for the particle following the trajectory $x^\mu(u)$ with a propagation period $T$:
\begin{eqnarray}
S[x^\mu(u); T]=
\frac{1}{2}\int_0^T  g_{\mu\nu} (x (u))\frac{dx^\mu}{d u}\frac{dx^\nu}{d u}  d u
\label{action}
\end{eqnarray}
For the evaluation of the  path integral we resort to the well known stationary phase approximation. In this approximation dominant contribution to the path integral comes from the  stationary points of the action, $S[x^\mu(u); T]$. These points are the solutions to the geodesic equation: 
\bea
\frac{d^2 x^{\lambda} }{d u ^2} + \Gamma^{\lambda}_{\mu\nu}\frac{dx^\mu}{d u}\frac{dx^\nu}{d u}=0
\label{eom1}
\eea
where $u$ is the affine evolution parameter.  Equation (\ref{eom1}) has a first integral: $\mathcal{L}=
\rm{constant}$.  This constant is fixed by also making a saddle point approximation to $T$ integral in (\ref{scalar}), giving:
\bea
\frac{m^2}{2} - \frac{\partial S}{\partial T}=0
\label{sta}
\eea
We identify the period satisfying the stationarity condition above as the classical (fundamental) period, which we denote by $T_{c}$. Using the Hamilton-Jacobi equation, we write the  stationarity condition in the configuration space as: 
\bea
m^2 + g_{\mu\nu} \frac{dx^\mu}{d u}\frac{dx^\nu}{d u} =0
\label{cons}
\eea
which identifies the normalization of the affine parameter: $u=\tau/m $, where $\tau$ is the proper time. The constraint given by (\ref{cons}) shows that the particle is restricted to move on a constant energy surface defined by: $\mathcal{E}=m^2$. With these observations the semiclassical form of the effective action can be given as:
\begin{equation}
\Gamma^{\rm scalar}_{\rm eff}\approx e^{i \frac{1}{2}m^2\, T_c} \int \sqrt{-g}\, \mathcal{P} \,e^{-iS[x, \, T_c]}  \,d^4x
\label{scalar1}
\end{equation}
Note that the integration cycle for the action has not been specified yet. With $S[x^\mu(u); T]$ evaluated on classical trajectory, all the terms arising from the second order contribution are combined in the prefactor:
\bea
\mathcal{P}=-i \frac{e^{-i m_p/2}}{T_c} \left (\sqrt{\text{det}\, \frac{\partial p_{\mu}(x(0))}{\partial x^{\nu}(T) }} \right)_{T=T_c}  \sqrt{\frac{2\pi}{\left| 
 \partial^2 \Delta / \partial T_c^2 \right|}}  , \quad \Delta =  \frac{1}{2} \left( i m^2 T_c -i S[x^\mu(u); T_c] \right)
\label{pre}
\eea
The Van-Vleck determinant and the last term above encode the fluctuations which are given by the variation of  the initial momenta with respect to the endpoints of the trajectory, ultimately giving the density of the trajectories emerging from the same neighborhood with different momenta. The Morse index, $m_p$ is the integer accounting for the number of negative eigenvalues of the determinant. Given the form of classical period satisfying the stationarity condition, the combined factors above  yield the momentum integrated spectrum of tunneling particles weighted by the classical action. To see this one may convert the path integral in (\ref{scalar}) to phase space path integral via Legendre transform and perform the momentum integrals  by the virtue of integrability.  In the next section we will articulate on this point by using the analytic results from the Schwinger effect.

\subsection{The need for Complex worldlines}

The worldline  approach was initially suggested for QED vacuum instability in constant electric field \cite{affleck}. In \cite{wl1}, the formalism was extended to inhomogeneous fields. The specification of  the closed orbits was discussed by Rubakov et. al \cite{rubakov}, where  the authors point out that classical trajectories living in the negative mass squared regions of the configuration space signal the tunneling instability. Recalling the scaling relation between $u$ and $\tau$, one may refer such trajectories as the tunneling orbits  parametrized by  imaginary affine parameter. For the deformation  $u \rightarrow i s $,  in non-derivative couplings, the  classical equations of motion acquire a sign change:
\begin{eqnarray}
\frac{d^2 x^\mu}{du^2}=-\frac{\partial V(x)}{\partial x_\mu} \quad \rightarrow\quad
\frac{d^2 x^\mu}{d s^2}=+\frac{\partial V(x)}{\partial x_\mu}
\label{scalar-imag}
\end{eqnarray}
where for the gauge coupling of the QED case, the deformation introduces a factor of "$i$" into the equations of motion.
\begin{eqnarray}
\frac{d^2 x^\mu}{du^2}=e\,F^{\mu\nu}(x)\frac{d x_\nu}{du} \quad \rightarrow\quad
\frac{d^2 x^\mu}{d s^2}=-i\,e\,F^{\mu\nu}(x)\frac{d x_\nu}{d s}
\label{qed-imag}
\end{eqnarray}
With the factor $i$ in front, above equations become complex so their solutions generically become complex as well. To illustrate this, in the following we briefly recall the use of complex worldlines in the Schwinger effect and show how prefactor contribution in (\ref{scalar1}) can equivalently be given by summing over the initial momenta of the orbits.  We will concentrate on the scalar QED, the generalization to spinor case is straightforward.

\subsubsection{The Schwinger effect}

Upon  deformation $u \rightarrow i s$  the stationarity condition in QED case becomes  $- \dot{x}_{\mu}(s) \dot{x}^{\mu}(s) + m^2=0$. Here $x^{\mu}(s)$ is the solution of (\ref{qed-imag}). The form of the effective action given by (\ref{scalar}) remains  the same except $g_{\mu\nu}$ is the Minkowski metric, (-,+,+,+) and  the covariant derivative represents the minimal coupling with the background gauge field: $\nabla_{\mu} = \partial_{\mu} - i q A_{\mu}$. The classical action, $S[x^\mu(s), \,T]$ is given by  
\begin{eqnarray}
S[x^\mu(s); T]=-i\int_0^T \left(\frac{1}{2}\frac{dx_\mu}{ds}\frac{dx^\mu}{ds} -i q \frac{dx_\mu}{ds}A^{\mu}(x(s))\right) ds
\label{actions}
\end{eqnarray}
Following \cite{wl1} closely, we will work with time dependent Sauter pulse given along $x_1$ direction: $A_1(x^0) = E_0/k \tanh{k x^{0}}$. For this potential  equations  of motion can be integrated for the vanishing momenta, yielding the solutions:
\bea
x^0_{\text{cl}}(s)&=&\frac{i}{k} \text{arcsin}\left[\frac{\gamma}{\sqrt{1+\gamma^2}} \sin{\left( \frac{ m k \sqrt{1+\gamma^2} s}{\gamma} \right)} \right] \nonumber\\
x^1_{\text{cl}}(s)&=&\frac{1}{k} \frac{1}{\sqrt{1+\gamma^2}}\text{arcsinh}\left[\gamma \cos{\left (\frac{ m k \sqrt{1+\gamma^2} s}{\gamma} \right)}\right]
\label{clt}
\eea
where $\gamma= m k /(q E_0)$ is the adiabaticity parameter and the classical period is simply:  $T_c=2 \pi \gamma /(m k \sqrt{1+\gamma^2}) $ . Using the above solutions the tunneling amplitude reads
\bea
\text{exp}\left[\Delta (x_{\text{cl}}, \, T_c)\right] = \text{exp}\left[-\frac{m^2 \pi}{q E_0}\left( \frac{2}{1+\sqrt{1+\gamma^2}}\right)\right]
\label{actc}
\eea
which reduces to Schwinger's result in the constant field limit, $\gamma \rightarrow 0$. Returning back to the prefactor expression (\ref{pre}), we see that the second order contribution arising from $T$ integral can be readily calculated once the exponent is specified in terms of the classical period, $T_c$. The remaining determinant factor depends on the zero modes of the secondary action, $\delta^2 S$. The calculation of zero modes has been carried out in \cite{wl1} and we give the details for integrating them out in the Appendix.  Here, we give the final result for the path integral, which upon  including the Morse index and the normalization factors,  leads to
\bea
\frac{\text{Im}\,\Gamma_{\rm eff}^{\rm scalar}}{V} = \frac{1}{8 \pi^3}  \frac{\left(1 + \gamma^2\right)^{5/4} \left (q  E_0\right)^{5/2}}{ m k} \text{exp}\left[-\frac{m^2 \pi}{q E_0}\left( \frac{2}{1+\sqrt{1+\gamma^2}}\right)\right]
\label{eff1}
\eea

\begin{figure}
\includegraphics[width=7.8cm,height=6cm]{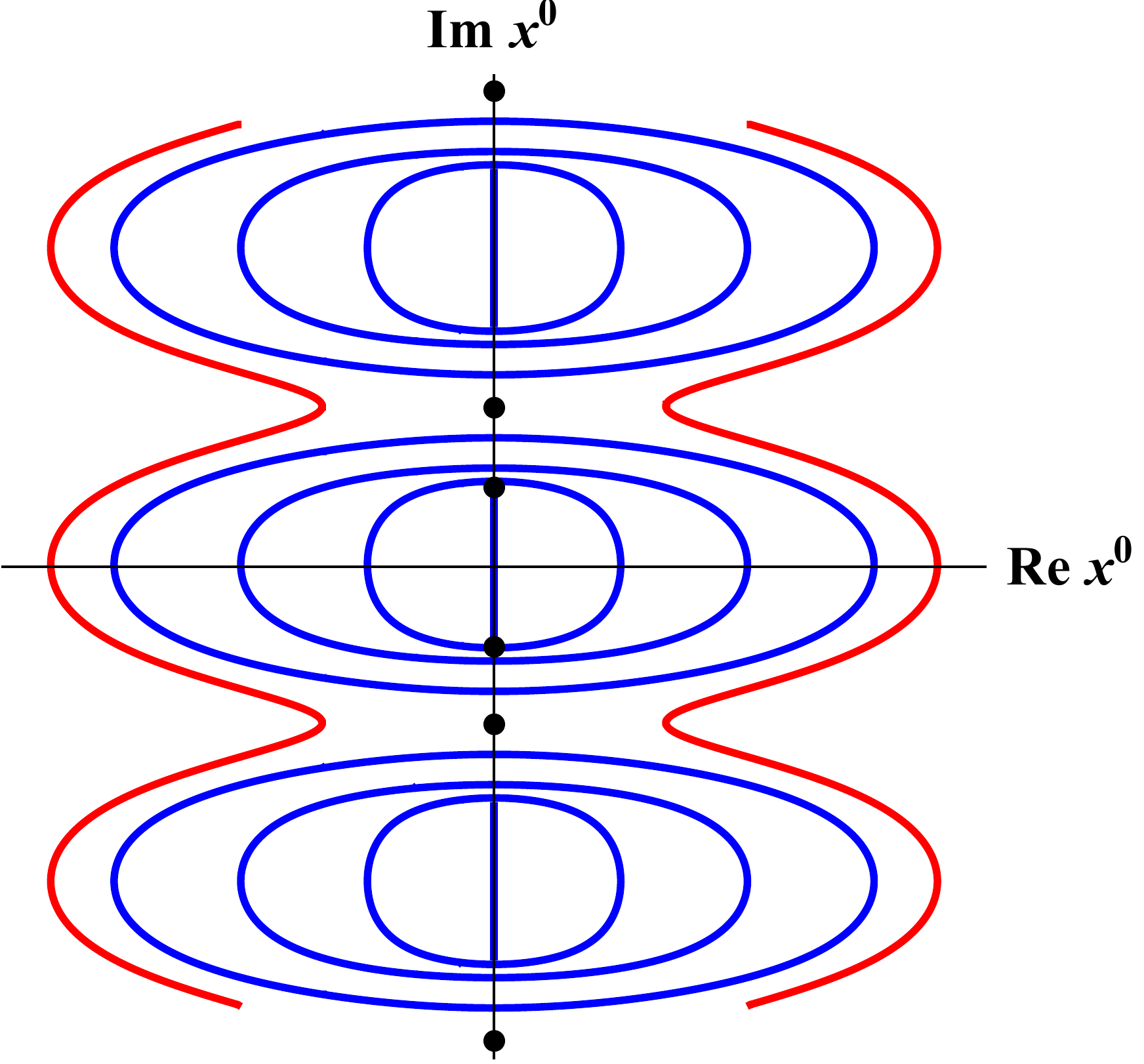}
\includegraphics[width=7.8cm,height=6cm]{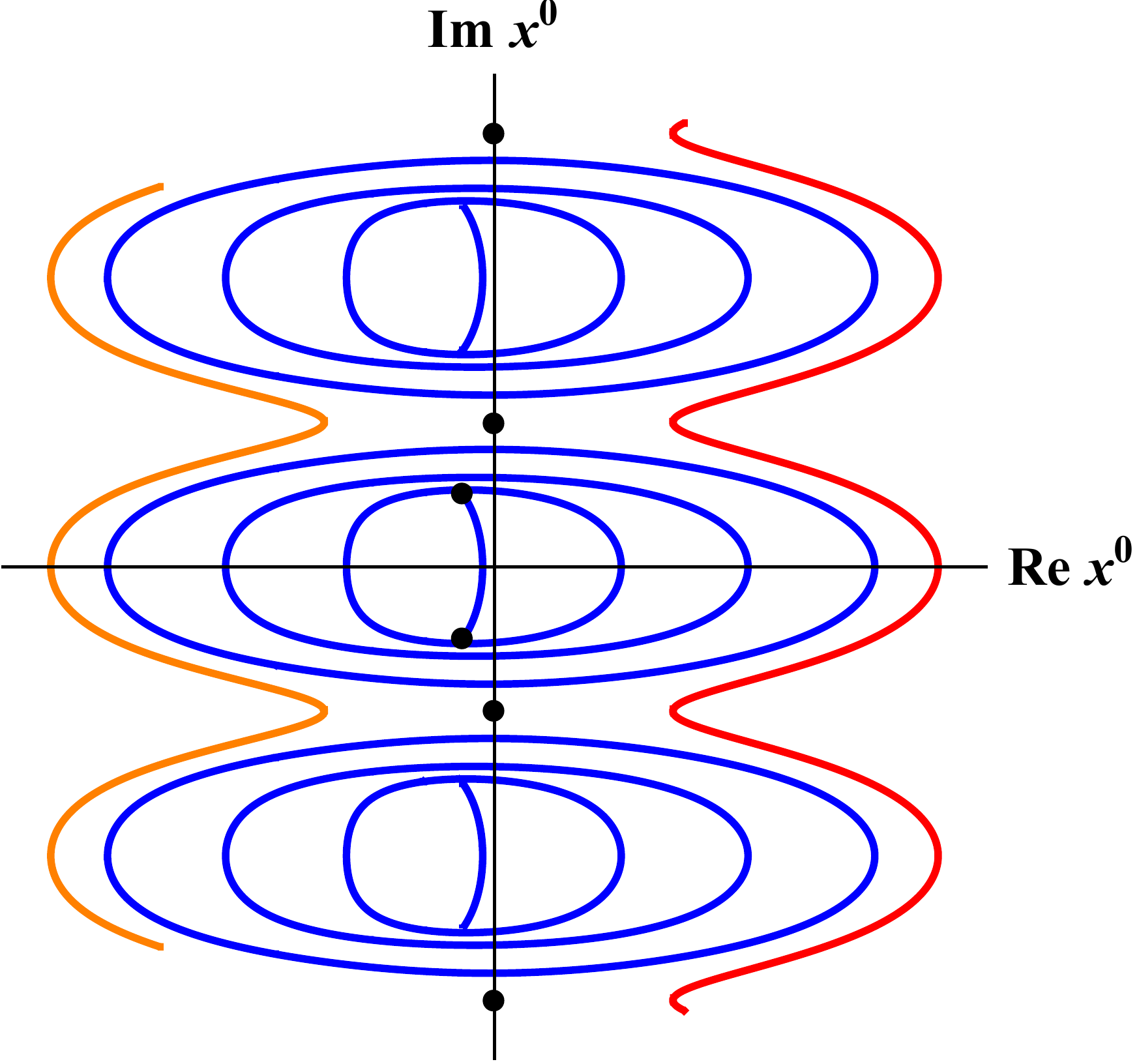}
\caption{Complex flow for the Sauter potential ($E_0=1/10, \, k=1/10, \, m=1, \text{and} \, f=1 $). On the left, the complex trajectories with vanishing conserved momenta form a symmetric flow with respect to the imaginary axis.  The classical solution given by (\ref{clt}) corresponds to the innermost orbit passing through the origin. The conjugate (critical) points are marked by the dots on the trajectory. The value of the action remains uniform on each trajectory marked by the same color. The families of orbits enveloped by the outermost trajectories are separated by the poles located at $x^0= i n \pi /k , \, n \in \mathbb{Z} $. On the right the longitudinal momentum is $p=1/10$, as a result the orbit flow gets deformed.}
\label{f1}
\end{figure}

 One may reach the above result  by integrating  the the tunneling amplitude over the initial momenta of  tunneling orbits. The contribution coming from zero modes are encoded by the momentum integrals. This is in fact expected because zero modes are associated with the continuous symmetries of the action.  To account for the initial momenta we convert the path integral into a phase-space path integral
\bea
\Gamma^{\rm scalar}_{\rm eff}=-i\int_0^{\infty}\frac{d T}{T} e^{i \frac{1}{2}m^2\, T} \int \sqrt{-g} \,d^4x\int\limits_{x(0)=x(T)} \mathcal{D}^{4}x \int  \mathcal{D}^{4}p \,e^{-i \int_0^T (p_{\mu} \frac{dx^\mu}{d u}-\mathcal{H}) d u}
\label{scalar2}
\eea
where Lagrangian and Hamiltonian densities are related by: $\mathcal{H}= p_{\mu}\frac{dx^\mu}{d u} - \mathcal{L}$. The Hamilton's characteristic function: $W[x^\mu(u); \, \mathcal{E}]= \int^{T}_{0}p_{\mu}\, \dot{x}^{\mu} d u$  is referred as the classical action of a particle moving on the constant energy surface  in the phase space. It is  related to the action via Legendre transform: $W[x^\mu(u); \mathcal{E}] =S[x^\mu(u); T] - \mathcal{E} T$, which immediately yields the relations: $\frac{\partial S}{\partial T}= \mathcal{E}$ and $\frac{\partial W}{\partial \mathcal{E}}=-T$. Noting that the gauge potential depends only on time, all the spatial components of the path integral above can be performed, producing delta functions over the spatial momenta. Thus, the momenta are conserved and the functional integrals over the conserved momenta reduce to ordinary momentum integrals: $\int \mathcal{D} p_{1} \mathcal{D} p_{2} \mathcal{D} p_{3} \rightarrow \frac{1}{8\pi^3}\int \ dp_1 dp_2 dp_3$. The stationary points of the phase space path integral satisfy
\bea
(e+ i f) \frac{d p_{\mu}}{d s} = -\frac{\partial\mathcal{H}}{\partial x^{\mu}}, \quad \, (e + i f) \frac{d x^{\mu}}{d s} = \frac{\partial\mathcal{H}}{\partial p_{\mu}}
\label{eom2}
\eea
where we have used $u \rightarrow s=(e + i f) u $. With $e$ and $f$ are real, such deformation reflects the reparametrization freedom of the affine parameter, in other words, we are free to choose the direction  in which we integrate the equations of motion in the complex plane. The corresponding integral motion, $\mathcal{H}=\text{constant}$ is fixed via the constant energy surface in the phase space:
 \bea
m^2 + g^{\mu\nu} p_{\mu}p_{\nu} =0
\label{cons2}
\eea
Note that when the external field is nonuniform, the period of the orbits on the constraint surface couples with the momenta: $T_c:=T_c(p_i, \,m)$ and because the characteristic function $W[x^0(s); \, m^2]$ depends only on the momenta, the correction terms brought by the deformation of the period and  the Van-Vleck determinant are  encoded by the momentum integrals, weighted by $e^{i W[x^\mu(s); \, m^2]}$.  Taking this into account and including the Morse index we may write:
\bea
\Gamma^{\rm scalar}_{\rm eff}\approx -i V \int  dp_1 dp_2 dp_3  \int\limits_{x(0)=x(T_c)} \sqrt{-g} \,\frac{d x^0} {T_c}  e^{-i m_p/2}  \,e^{-i \int_0^{T_c} p_{0} \frac{dx^0}{d s} d s}
\label{scalar3}
\eea
Here, we left  $T_c$ in the exponent to emphasize the fact that characteristic function is to be evaluated on the phase space orbit over the full closure period. Periodic orbits of (\ref{eom2}) can be found upon appropriately specifying the initial conditions. Bounded orbits are generically located in the vicinity of  the critical points, where the one form $p_0\, dx^0\equiv p_0 \,\dot{x}^0 ds $ vanishes. In one dimensional problems, the locations of the critical points can  easily be extracted  from (\ref{cons2}) by specifying $p_0$ as a function of the conserved momenta and the external parameters. Here, for the time dependent potentials critical points are  located on the complex $x^0$ plane, generally in the form of complex conjugate pairs. The dominant contribution to path integral in the weak field limit, $E_0\ll m^2$ and $\gamma \ll 1$ is given by the complex trajectory lying in the neighborhood of the critical point pair, which is closest to the real axis (see Figure \ref{f1}). 

Before proceeding with the momentum integrals, the remaining piece of (\ref{scalar3}) to look at is the integral over the initial points of the trajectory. Generically, all the points located on the closed trajectory contribute to the volume integral. This simply reflects the freedom in  choosing the starting point on the trajectory when evaluating the action. It is obvious that $W[x^{\mu}(s); \, m^2]$ does not depend on the particular choice of $x^{\mu}(0)$ since  it remains invariant under proper time translations along the trajectory. But due to this translational freedom, the volume element contains a multiplicity factor given by the path length of the trajectory. This multiplicity factor cancels against  $1/T_c$ factor appearing in (\ref{scalar3}), therefore normalizing the volume integral. As a result the path integral over a single trajectory becomes
\bea
\Gamma^{\rm scalar}_{\rm eff}\approx -i \frac{V}{8\pi^3} \int  dp_1 dp_2 dp_3  \, e^{-i m_p/2}  \,e^{-i \int_0^{T_c} p_{0} \frac{dx^0}{d s} d s}
\label{scalar4}
\eea
To show that above expression indeed leads to (\ref{eff1}), the remaining task is to specify the characteristic function on the classical orbits with non vanishing momenta. The analytical form of the orbits in (\ref{clt}) for arbitrary momenta is unknown but the form of the characteristic function evaluated over the closure period can be deduced by noting the equivalent representation of the effective action \cite{niki}
\bea
\text{Im} \, \Gamma^{\rm scalar}_{\rm eff}= -\frac{V}{8\pi^3} \int d^3 p \, \text{log}\left( 1 - w (p) \right)
\eea
where $w(p)$ is the backwards scattering amplitude, obtained by solving for the asymptotic solutions of 
\bea
\partial^2_0\varphi (x^0) + p_0^2 (x^0) \varphi =0 , \quad p_0 = \sqrt{m^2 + p^2_2 + p^2_3 + \left (p_1- q A_1 (x^0) \right)^2}
\eea 
The amplitude $\varphi (x^0)$ is the temporal component of the wavefunction of a scalar particle propagating in a  time dependent electric field. The spatial components are just the free particle solutions due to the translational invariance and have been integrated out. The initial and final vacuum states are respectively given by the boundary conditions
\bea
\lim_{x^{0} \to -\infty} \varphi (x^0)   \rightarrow e^{i p^{-}_0} &,&   \quad \lim_{x^{0} \to -\infty} p_{0}(x^0)  = p^{-}_0\nonumber\\
\lim_{x^{0} \to \infty} \varphi (x^0)     \rightarrow c_1 (p) e^{i p^{+}_0} + c_2 (p) e^{-i p^{+}_0} &,& \quad \lim_{x^{0} \to \infty} p_{0}(x^0)  = p^{+}_0
\eea
with the unitarity condition, $\left | c_1(p) \right|^2 - \left| c_2(p) \right|^2 =1 $. The above scattering problem for the Sauter potential is exactly solvable. The momentum dependence of the scattering amplitude is given by
\bea
w(p)= \frac{ \left| c_2(p) \right|^2}{\left| c_1(p) \right|^2} = \frac{\cosh{\frac{\pi}{k}\left(i \kappa + p^{-}_0 -p^{+}_0 \right)}\cosh{\frac{\pi}{k}\left(i \kappa - p^{-}_0 +p^{+}_0 \right)}}{\cosh{\frac{\pi}{k}\left(i \kappa + p^{-}_0 +p^{+}_0 \right)}\cosh{\frac{\pi}{k}\left(i \kappa - p^{-}_0 -p^{+}_0 \right)}}
\eea 
where $\kappa= \sqrt{k^2 - 4 q^2 E_0^2/k^2}$. In the weak field limit and for  $\gamma \ll 1$, the scattering amplitude becomes \cite{popov}:
\bea
w(p)\approx \text{exp}\left[ - \frac{\pi}{k} \left(p^{+}_0 +p^{-}_0 - \frac{2 q E0}{k}  \right) \right]
\label{bsa2}
\eea 
The exponent above is  precisely the Hamilton's characteristic function evaluated on the classical trajectory with non-vanishing momenta. This could be verified by direct numerical integration or computing the  characteristic function as a Cauchy integral with the contour chosen as the classical trajectory. We may now evaluate (\ref{scalar4}) by using (\ref{bsa2}) and noting that the Morse index for the classical trajectory is 2 (see Appendix).  By expanding the exponent in the limit $p_i/m \rightarrow 0$ and up to quadratic order in momenta, the path summation turns into  Gaussian integrals
\bea
\Gamma^{\rm scalar}_{\rm eff}\approx i \frac{V}{8\pi^3}  \text{exp}\left[-\frac{m^2 \pi}{q E_0}\left( \frac{2}{1+\sqrt{1+\gamma^2}}\right)\right] \int  dp_1 \text{exp}\left[- p^2_1\frac{\pi \gamma^3}{m k \left(1+\gamma^2\right)^{3/2}}\right] \,  
\left( \int  dp \, \text{exp}\left[- p^2\frac{\pi \gamma}{m k \left(1+\gamma^2\right)}\right]\right)^2
\label{scalar5}
\eea
The resultant integrals over the transverse momenta produces the free particle prefactor, $1/2 \pi T_c$. The remaining integral over $p_1$ encodes the collective contribution of Van-Vleck determinant in $x^0-x^1$ plane and also the leading term coming from the $T$ integral. By performing the integration over $p_1$,  it is easy to see that (\ref{scalar5}) indeed matches with (\ref{eff1}). 

\subsubsection{Hawking Radiation} 

Returning to our case in curved background we see that it is favorable to perform the path summation over the phase space. Owing to the axisymmetry of the Kerr metric the spatial path integrals over $t$ and $\phi$ in (\ref{scalar2}) can be done, producing the delta functions for the momenta, $p_t$ and $p_{\phi}$. Thus, $-p_t=\omega$ and  $p_{\phi}=j$ are conserved and functional integrals over conserved momenta reduce to ordinary momentum integrals as before:  $\int \mathcal{D} p_{\phi} \mathcal{D} p_{t} \rightarrow \int d\omega\ d j$ \cite{dewittc}. Using the WKB ansatz on the remaining phase space path integrals, the imaginary part of the vacuum to vacuum transition amplitude reads (Henceforth we drop the subscript on $T_c$)
\begin{equation}
\text{Im}\,\Gamma^{\rm scalar}_{\rm eff} \approx  \mathcal{N}\int d\omega \, d j  d p_{\theta}\int \sqrt{-g} \,d^4x   \,e^{-i \int_0^{T} p_{r}  \dot{r} d u } \,e^{-i \int_0^{T} p_{\theta}  \dot{\theta} d u}
\label{scalar6}
\end{equation}
where we have absorbed the phase space normalization and  the sign factors arising from the Morse index into the prefactor $\mathcal{N}$.  The appearance of the measure, $dp_{\theta}$  is necessary because path summation is  to be performed over the initial values of $p_{\theta}$ resulting in the bounded motion. In addition, it  ensures that the phase space volume has the correct form that is preserved by the Hamiltonian flow.  To evaluate  (\ref{scalar6}), we need to specify the integration cycles for the action, which  in turn should give the tunneling exponent over the horizon area. As  we will demonstrate,  on the integration cycles  Hamilton's characteristic function does not  depend on the value of $p_{\theta}$. This has an important consequence. It is well known that Kerr metric is classically an integrable system, where there are as many constants of motion in involution as the number of the coordinates.  In the usual phase space coordinates the integrability is reflected in the existence of Carter's constant, which can be stated as a constraint between the momenta and $\theta$:
\bea
\mathcal{C}=p^2_{\theta} + \cos^2{\theta} \left(-a^2 (\omega^2 - m^2) + \frac{j^2}{\sin^2{\theta}}\right)
\label{carter}
\eea
where $a$ is the rotation parameter of black hole. The important message here is the following. By observing the $p_{\theta}$ independence of the tunneling exponent and using the above constraint,  $p_{\theta}$ in (\ref{scalar6}) can be integrated out when summing over a family of tunneling cycles, for which the value of $\mathcal{C}$ remains fixed. We will use this fact later in order to give an analytic expression for the emittance of  black hole at the event horizon:
\begin{equation}
\sigma \approx  \mathcal{N} 4 \pi M r_+ \int d\omega \, d j  d p_{\theta}\int  \,d \theta \sin{\theta}    \,e^{-i \int_0^{T} p_{r}  \dot{r} d u } \,e^{-i \int_0^{T} p_{\theta}  \dot{\theta} d u}
\label{emit}
\end{equation}
which we define as the total tunneling rate at the outer horizon per unit time. Here, the extra factors come from the reduced metric determinant and the angular integration.  It is important to note that the path summation above does not take into account the greybody contribution. Tunneling flux by definition accounts for the density of  the emitted particles at the vicinity of the horizon. The further damping of the tunneling amplitude by the gravitational potential must be accounted by the additional cycles which extend to the region far away from  the horizon. Formally, this means  (\ref{emit})  must include the radial coordinate,  extending from $r_+$ all the way to the infinity.  Another subtle point is that the flux in (\ref{emit}) is defined in the continuum limit.  In the standard computations of the flux all the modes except the energy is quantized via the eigenvalues of the separation constant, $\mathcal{\lambda}_{\,l\, j}$   that appears in the wave equation \cite{page2, churilov,brito}. Heuristically speaking, the continuum limit of angular momentum modes here is encoded by the Carter's constant  such that: $\mathcal{\lambda}_{\,l\, j} \xrightarrow{\text{continuum}} \mathcal{C} + j^2$ \cite{brill}.  

Where are the tunneling orbits ? The expression (\ref{emit}) makes the tacit assumption that the initial points of the tunneling orbits  must lie arbitrarily close to the event horizon.  To see this, it is illustrative to consider the radial action in (\ref{emit}) as   a Cauchy integral. The tunneling exponent is then given by the residues of the classical action at the event horizons. Recalling the Hamilton-Jacobi formalism, the integration contour for the residue is adopted as a small circle in the complex $r$ plane, encircling the pole at the horizon. In the worldline framework, such a choice is not given as a priori, rather it is constructed by using the closed orbits from the Hamiltonian flow.  But the construction is not arbitrary; as we will explicitly show the integration cycles giving the tunneling exponent precisely correspond to the paths on which one performs the analytic continuation of the modes across the horizon. In view of this fact, tunneling exponent in fact emerges as a relative phase difference across the event horizon,  as illustrated by the Damour-Ruffini method\cite{dr}.  In order to reach this result, we must first  fix the domain of integration cycles by using the constant energy surface defined by
\bea
\frac{m^2}{\left(c+i d\right)^2} + g^{\mu\nu} p_{\mu} p_{\nu} =0
\label{cons3}
\eea
The above equation, depending on the values of $c$ and $d$, fixes the location of critical points where the one-form $p^{1,\,2}_r \dot{r} \, d s \equiv p^{1,\,2}_r \, dr $ vanishes. Here $p^{1,\,2}_r$ denote the algebraic solutions of (\ref{cons2}).  The important observation is that bounded orbits of the Hamiltonian reside in the vicinity of the critical points therefore the relative location of critical points with respect to event horizons is crucial. For $c=0$ and $d \in  \mathbb{R}$, the two of the critical points of $p_r$  are generically located on the real axis, in between the event horizons. This configuration supports bounded complex orbits which smoothly fill the regions located in between and outside the horizons. As we will demonstrate, the phase difference across the horizons is encoded by these orbits.  To find the orbits, we fix $e=0$ and the magnitude of $f$ to unity in (\ref{eom2}). The sign of $f$ will be specified depending on whether the trajectory starts inside or outside the horizon.

\subsection{Hamiltonian Flow for Complex worldlines}

\textbf{The choice of coordinates:} Spheroidal coordinates, in which the axisymmetry of Kerr geometry becomes apparent, is convenient for the integration of (\ref{eom2}). These include Boyer-Lindquist (BL), Kerr, Painleve-Kerr\cite{pkc} and Doran coordinate systems\cite{dc}. The Hamiltonian flow in these coordinate systems form a smooth family of nested integral curves, encompassing the inner and outer horizon regions and they are almost identical to one other. The findings reported here are based on BL and Kerr coordinates which are computationally less demanding, but the identical results for the tunneling amplitude can easily be generated for any other choice of spheroidal coordinates.   We begin by analyzing the flow of radial geodesics.  For this we set: $p_{\theta}(0)=0,\, \theta(0)=\pi/2 $. The initial condition for radial momentum is given by the solutions, $p^{1, 2}_r$. 

\subsubsection{BL Coordinates}

\begin{figure}
\includegraphics[width=8.1cm,height=5.5cm]{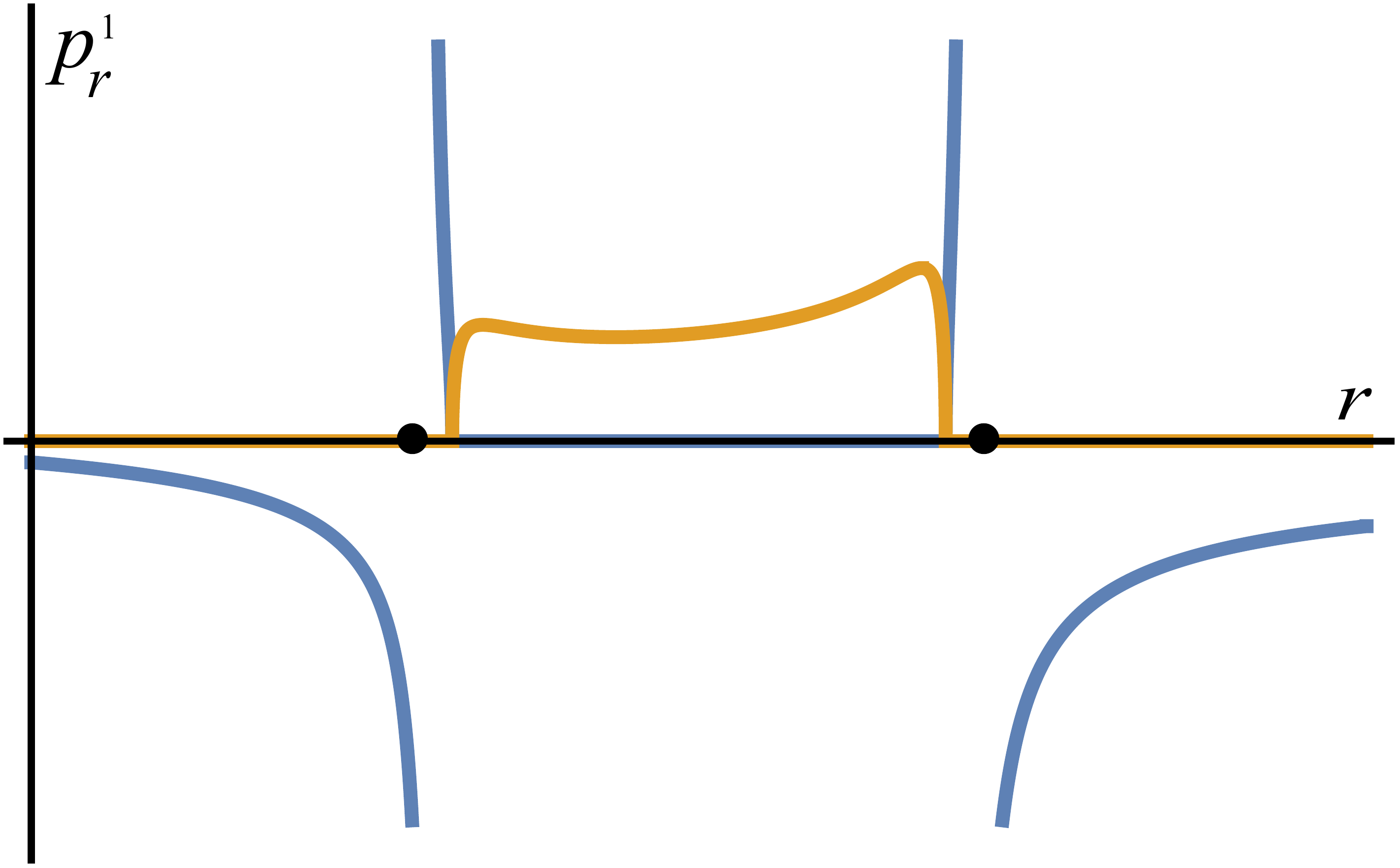}
\includegraphics[width=8.1cm,height=5.5cm]{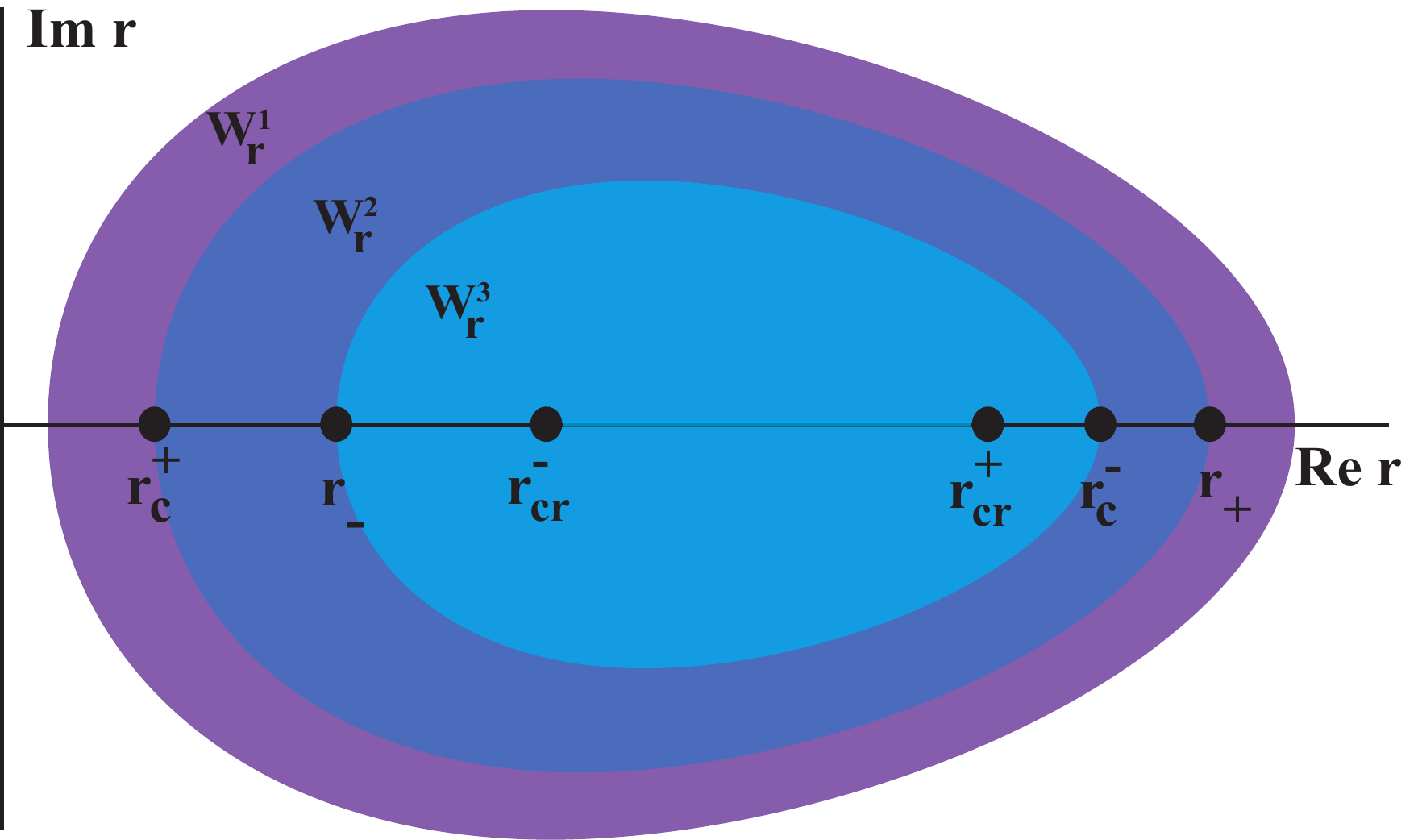}
\caption{The radial momentum with real (blue) and imaginary (yellow) parts (left). The dots indicate the location of event horizons. Note that the imaginary part vanishes at the critical points (see the text). The parameters are chosen as: $M=10^5, \, a=9.7 \times 10^4,\, m=1, \, \omega =5\times 10^{-2}, \, j=10^{-2}, \, \theta(0)=\pi/2, \, p_{\theta}(0)=0, \, \text{and}\, \, d=1$. On the right is the schematic representation of the Hamiltonian flow formed by complex geodesics. The geodesics cover an elliptical area, where each shaded region designates the family of the worldlines for which  the value of $W_{r}$ remains uniform. The discontinuities in the classical action yield the tunneling exponents for the inner and outer horizon and in BL coordinates they occur when passing across the event horizons and the conjugate points.} 
\label{f2}
\end{figure}

The radial dependence of $p^1_r$ in BL coordinates is plotted in Figure (\ref{f2}). The other solution, $p^2_r$ just differs by a minus sign: $p^1_r=-p^2_r$. The poles of $p^{1}_r$ are located at the event horizons:  $ r_{\pm} = M \pm \sqrt{M^2-a^2}$.  As discussed earlier, the points  where $p_r \dot{r} $ becomes zero are the critical points which here coincide with the zeros of $p^{1}_r$. The locations of the poles and the critical points are important for the  Hamiltonian flow: poles repel the neighboring trajectories, whereas critical points  behave as attractors.  The reason that geodesic flow remains almost identical for the choice of BL, Kerr, Doran and Painleve-Kerr coordinates is because in these coordinates the distribution and the number of poles and zeros remains  the same. In spheroidal coordinates, there are four critical points on the complex plane. For fixed $a$ and $M$ and also for fixed mass $m/d$, the location of the critical points vary as a function momenta and $\theta$.  Two of these points  remain in the negative $r$ plane, on the real axis or appear as complex conjugate of one other. For  $c=0$ and $d \in  \mathbb{R}$,  the remaining pair of critical points, $r^{\pm}_{\text{cr}}$ are generically located in between the two event horizons, on the real axis or as complex conjugates. Exception to this comes from the extremely high momenta region where the energy of the tunneling particle by far exceeds the chosen mass scale, $m/d$ or where the angular momentum takes values in the deep superradiant regime.  Here, if the two pairs of critical points are separated well enough, the latter pair of critical points, $r^{\pm}_{\text{cr}}$, gives rise to bounded orbits encoding phase discontinuities across the event horizons.

To see this,  one may fix the initial momenta and obtain the geodesic flow by varying $r(0)$ on the real axis, from inner horizon to outer horizon region. Figure (\ref{f2}) shows the geodesic congruence belonging to $r(s)$ in the complex domain. In the subregions denoted by the shaded areas the value of the action, $W_r = -i\int _0^T p_r \dot{r}\, ds $ remains uniform. While passing the event horizons,  the value of $W_r$ makes a jump because of the phase difference brought by the pole of $p_r$. The same phase difference also occurs for the points $r^{\pm}_{\rm c}$, which are conjugate to event horizons, $r_{\pm}$. Conjugate points mark the initial points for which radial component of the trajectory over the half period coincides with the horizon: $r^{\pm}_{\rm c}(T/2) = r_{\pm}$. For instance, if $r(0)$ approaches $r^{+}_{\rm c}$ from the left (right), trajectory will cross the real axis just outside (inside) the outer horizon. The  radial momentum, $p^1_r$  over the half period is mapped into the other solution $p^2_r$ (and vice versa): $p^1_r(T/2, r(T/2))=p^2_r(0, r(T/2))$. Because of this, the values of $p^1_r(s, r(s))$ across the initial points: $r(0)=r^{\pm}_{\rm c} \pm \epsilon\, (\epsilon\ll 1) $ are separated by the poles of $p^2_r$  over the half period, therefore $W_r$ converges to different values across the conjugate points as well.

\textbf{Integration cycles:} The tunneling exponent is given by the total phase difference across the horizons.  By looking at the Hamiltonian flow, the phase difference can readily be seen to be $W^1_r-W^2_r$ at the outer horizon. Formally one may obtain this result by combining the trajectories located on both sides of the outer horizon and use them as integration contours for the action. The combined integration cycle consists of the trajectory, $r(s)$ located just outside the horizon with $s=i u, \, (f=1)$ and the trajectory just inside the horizon with  $s=-i u , \, (f=-1)$. The classical action evaluated on this cycle in effect gives the residue located at $r_+$.  We should however mention that  $W^1_r-W^2_r$  corresponds to the tunneling exponent, apart from the fact that it gives half of the correct result.  This is related to the factor two ambiguity that arises in several methods (see \cite{ahmedov} and the references therein). Here in the worldline approach it only shows up in BL coordinates but can be resolved by the following observation. Note that in the beginning we defined the tunneling exponent as the total phase difference, which is in fact brought by the poles of the solutions $p^{1,2}_{r}$. In BL coordinates both solutions have poles located at the horizons, thus the contribution coming from the cycle where $p_r(0)=p^2_r$ should also add up to this phase difference.  As a result, the correct tunneling exponent is indeed given by $2W^1_r- 2W^2_r$. The tunneling amplitude is accordingly given by 
\bea
e^{ 2W^1_r- 2W^2_r}&=&e^{-\left| \omega-j \Omega_{+}\right| / \, T^{+}_{H}} , \qquad  W^1_r- W^2_r  <0  
\label{tun}
\eea
where the angular velocity, $\Omega_+$  and the temperature, $T^{+}_H$ of the outer event horizon are defined as
 \bea
\Omega_{+}&=& \frac{a}{a^2+ r^2_{+}},\quad T^{+}_{H}=\frac{r_+-r_-}{4\pi(r_{+}^2+a^2)}\nonumber
\label{antemp}
\eea
The tunneling exponent in (\ref{tun}) is given as an absolute value for the following reason: as one keeps $\omega$ constant and increases $j$,  the critical point $r^+_{\rm cr}$ moves towards $r_+$  and at the onset of superradiance ($j=\omega/ \Omega_+$), $r^+_{\rm cr}$ coincides with the horizon $r_+$. From this point on, further increase in angular momentum makes $r^+_{\rm cr}$  to move back  from $r_+$ towards $r < r_+$. The crucial point is that when passing into superradiant regime $W_r$ does not reverse the sign on the cycle, in other words tunneling exponent still remains negative: $2W^1_r- 2W^2_r= -\left|\omega - j \Omega_+ \right|/T^+_H$.  It is obvious that for the modes, $\omega  > j \Omega_+ $ the emission probability is exponentially suppressed. The weight factor is given by the repeated traversals of the same trajectory: 
\bea
\sum^{\infty}_{n=1}e^{- \frac{n}{T^{+}_H} (\omega- j \Omega_+)}=\frac{1}{e^{(\omega- j \Omega_+)/T^+_H}-1}
\label{weight}
\eea
which is neglected in the path summation for brevity. For superradiant modes if the tunnelling exponent indeed becomes positive, the expansion in (\ref{weight}) does not converge and we would have a difficulty in interpreting (\ref{emit}) as a finite sum. In fact, with the weight factor chosen as $1/(e^{(\omega- j \Omega_+)/T^+_H}-1)$,  there is nothing to stop momentum integrals in  (\ref{emit}) from growing quartically, when $\omega < \Omega_+ j$. In the remainder of this analysis  we assume tunneling exponent remains negative in the superradiant regime.

The integration cycle for the inner horizon is given in a similar manner. The only difference is that on the trajectory lying just outside the inner horizon we choose $s=i u$, while on the inner trajectory we have $s=-i u$, so that  $W^2_r- W^3_r  <0$. This leads to:
\bea
e^{ 2W^2_r- 2W^3_r}&=&e^{-\left| \omega-j \Omega_{-}\right| / \, T^{-}_{H}} \nonumber  \\ 
\Omega_{-} &=&  \frac{a}{a^2+ r^2_{-}},\quad T^{-}_{H}=\frac{r_+-r_-}{4\pi(r_{-}^2+a^2)}  
\label{tuni}
\eea
Note that the value of $W_r$ on both cycles is reparametrization invariant because it is given as a residue of the one form.

\subsubsection{Kerr Coordinates}

\begin{figure}
\includegraphics[width=8.1cm,height=5.5cm]{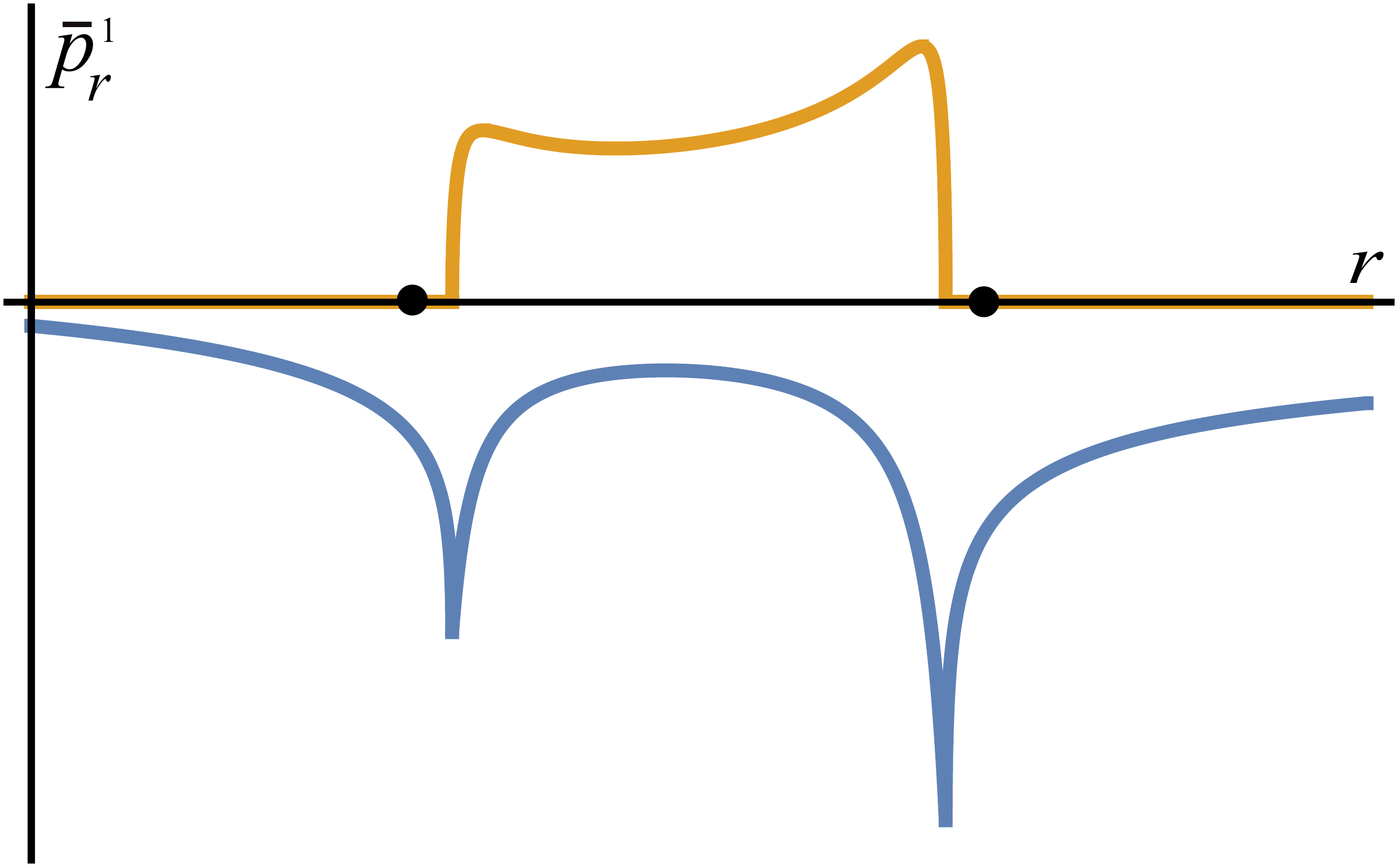}
\includegraphics[width=8.1cm,height=5.5cm]{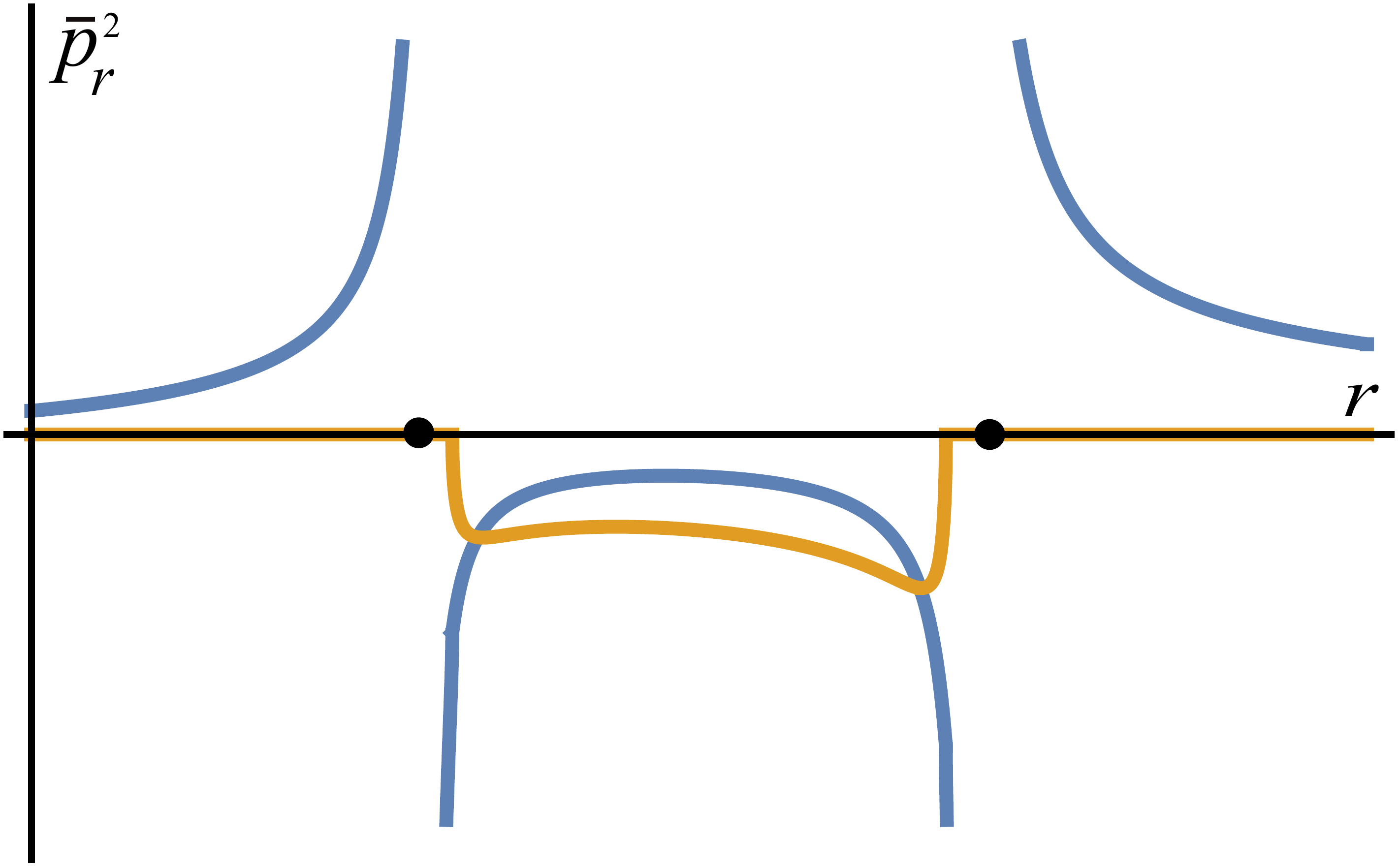}
\caption{The radial dependence of the solutions $\bar{p}^{\,1, \,2}_{r}$ in Kerr coordinates with real (blue) and imaginary (blue) parts. The first solution on the left stays regular at the horizons. The parameters are chosen as: $M=10^5, \, a=9.7 \times 10^4,\, m=1, \, \omega =5\times 10^{-2}, \, j=10^{-2}, \, \theta(0)=\pi/2 , \, p_{\theta}(0)=0 \, \text{and}\, \, d=1$}
\label{f3}
\end{figure}

The passage from BL coordinates to Kerr coordinates is given via singular coordinate transformation:
\bea
d\bar{t} = dt + \frac{a^2 + r^2}{a^2-2M r + r^2} dr, \qquad d \bar{\phi} = d \phi + \frac{a}{a^2-2M r + r^2} dr
\label{tra}
\eea
This transformation maps  $p^{1, \,2}_{r}$ in BL coordinates to $\bar{p}^{\,1, \,2}_{r}$ in Kerr coordinates  and leaves the remaining momenta unchanged. Figure (\ref{f3}) shows the radial dependence of  $\bar{p}^{\,1, \,2}_{r}$: the first solution, $\bar{p}^{\,1}_{r}$ stays regular at the horizons, whereas $\bar{p}^{\,2}_{r}$ remains singular. As a result, the value of the action $\bar{W}_r$ remains uniform across the horizons when $\bar{p}_r =\bar{p}^{\, 1}_r $, and it remains uniform across the conjugate points  when  $\bar{p}_r =\bar{p}^2_r $. Therefore, the phase difference can be obtained by only considering the trajectories, whose initial momentum is given as $p_r(0)=\bar{p}^2_r$. Note that these observations remain valid in Doran and Painleve-Kerr coordinates as well.

The  integration cycle is given in a similar fashion as before: the trajectory to the right of the outer horizon flows clockwise ($s=i u$)  whereas the inner trajectory flows anticlockwise ($s=-i u$). The resultant contour encircles the outer horizon clockwise and yields the value of the classical action as: $\bar{W}^1_r- \bar{W}^2_r$. The integration cycle for the inner horizon is chosen similarly. The only difference is that   the trajectory flows anticlockwise ($s=-i u$) outside the inner horizon  whereas the inner trajectory flows clockwise ($s=i u$). This cycle encircles the horizon clockwise as before and yields, $\bar{W}^3_r- \bar{W}^2_r$.
Accordingly, tunneling amplitude is given by:
\bea
e^{\bar{W}^1_r- \bar{W}^2_r}&=&e^{-\left| \omega-j \Omega_{+}\right|/ \, T^{+}_{H}} , \qquad  \bar{W}^1_r- \bar{W}^2_r < 0  \nonumber\\
e^{\bar{W}^3_r-\bar{W}^2_r}&=&e^{-\left| \omega-j \Omega_{-}\right| / \,T^{-}_{H}}  , \qquad \bar{W}^3_r- \bar{W}^2_r  < 0 
\label{tun2}
\eea
Note that the factor of two problem in regular coordinates disappears. All the relevant phase change is encoded by the poles of $\bar{p}^{\, 2}_r$ and in fact, the coordinate transformation in (\ref{tra}) doubles the value of the radial momentum at both horizons: ${\rm lim}_{r \rightarrow r_{\pm}}| \bar{p}^{\, 2}_r / p^{1,\,2}_r | = 2$.  

\begin{figure}
\includegraphics[width=8.1cm,height=5.5cm]{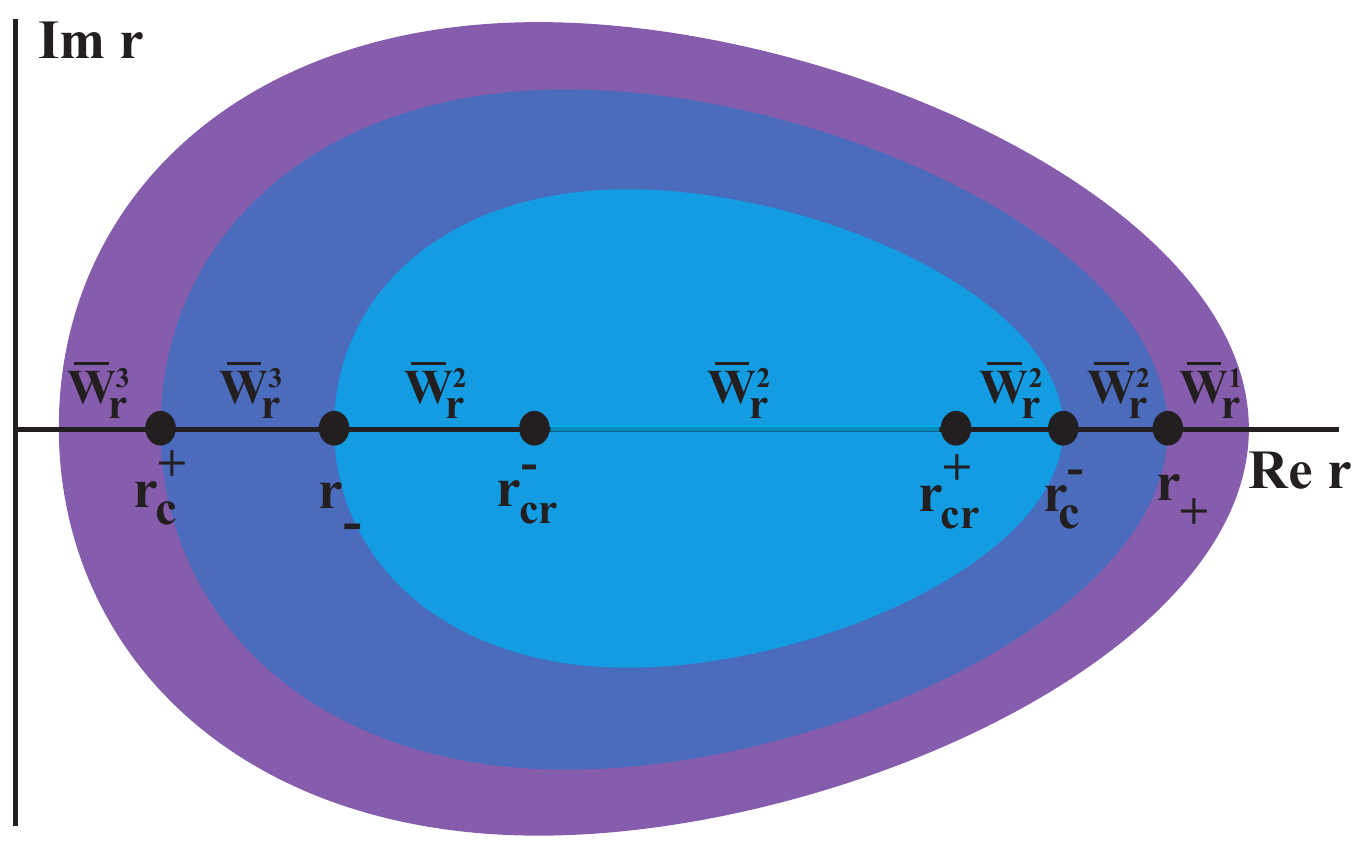}
\caption{Hamiltonian flow in Kerr coordinates with $p_{r}(0) =\bar{p}^{2}_{r} $. As in BL coordinates, the shaded areas are composed of  radial geodesics lying in the regions designated by $r_{\pm}$ and $r^{\pm}_{\text{c}}$.  Unlike in BL coordinates, the discontinuities in the action either occur  across the horizons(when $p_{r}(0) =\bar{p}^{2}_{r} $) or across the conjugate points(when $p_{r}(0) =\bar{p}^{1}_{r}$).} 
\label{f4}
\end{figure}

\textbf{Analytic continuation:}  Here we show that the integration cycles specified above are in fact the key ingredients in analytic continuation of the modes across the horizons. We believe establishing this connection is particularly illustrative for putting the complex worldline picture together with the approach of Damour Ruffini and also with the path integral derivation of black hole radiance given by Hartle and Hawking, who have also made use of the complex paths parametrized by imaginary affine parameter \cite{hh}.  To start, we consider the overlap of the modes located just to the left and to the right of the conjugate point: 
\bea
\braket{\psi(r^{+}_{\rm c}-\epsilon) |\psi(r^{+}_{\rm c}+\epsilon)}\approx \text{exp} \left (-i  \int^{r^{+}_{\rm c}+\epsilon}_{r^{+}_{\rm c}-\epsilon} \hspace*{-0.5cm}p_r\, dr \right) 
\label{in}
\eea
where one may choose the integration contour as an infinitesimal arc joining the left and right neighborhoods of $r^+_c$ by a suitable choice of $e$ and $f$. On such a contour the magnitude of the above amplitude can taken as unity since radial momentum remains  regular. The basic premise here is that  $\psi(r^{+}_{\rm c}+\epsilon)$ can be rewritten as the analytical continuation of the mode located just inside the outer horizon:
\bea
\ket{\psi(r^{+}_{\rm c}+\epsilon)}=\text{exp} \left (-i  \int^{T/2}_0 \hspace*{-0.5cm} p_r \, \dot{r} \, ds \right)_{-} \ket{\psi(r_{+}-\epsilon)}
\label{ac1}
\eea 
Here the subscript $-$ shows  that the integration path is chosen as the complex orbit, starting at the left of the outer horizon and ending at $r^{+}_{\rm c}+\epsilon$  (see Figure \ref{f4}). Similarly we have:  
\bea
\bra{\psi(r^{+}_{\rm c}-\epsilon)}=\bra{\psi(r_{+}+\epsilon)}\text{exp} \left (i  \int^{T/2}_0 \hspace*{-0.5cm} p_r \, \dot{r} \, ds \right)_{+} 
\label{ac2}
\eea 
where we have used the orbit starting just outside the horizon and ending at $r^{+}_{\rm c}-\epsilon$. To determine the direction of the flow, the sign of $f$  must be chosen in accordance with the exponential dominancy of the solutions and the  sign convention employed. For this, we first note that the amplitude of the modes populated inside the horizon are exponentially larger than the amplitude tunneling out and because of this $\braket{\psi(r_{+}+\epsilon) |\psi(r_{+}-\epsilon)}$ is exponentially small (subdominant). Taking this into account and recalling that  the amplitude in (\ref{in}) is order of unity,  the combined phase factor coming from (\ref{ac1}) and (\ref{ac2}) must be exponentially large (dominant), in other words,  the resultant exponent must be positive. Note that the total phase over the full period was previously given as $\tilde{W}^1_r- \tilde{W}^2_r $ which is negative for $f=1$. Here, the sign in front of the phases are reversed so in order for the total exponent be positive we should take $f$ as 1. This corresponds to choosing the integration paths as the lower segments of the complex orbits, oriented clockwise.  In computing the conjugate amplitude, $\braket{\psi(r^{+}_{\rm c}+\epsilon) |\psi(r^{+}_{\rm c}-\epsilon)}$, the signs of the exponents are reversed, therefore we should pick $f=-1$ for the exponential dominancy. Accordingly, the integration paths are anticlockwise and given by the upper segments of the orbits. Here we can make the equivalent choice and fix $f=1$ such that flow is given by the upper segments but clockwise. This way the resultant phase factor is given in terms of  the inner and outer orbits with a full period. In view of these observations we can express the flux density at the outer horizon as
\bea
\left| \braket{\psi(r_+ - \epsilon ) |\psi(r_+ + \epsilon)}\right |^2 \text{exp} \left (i  \int^{T}_0 \hspace*{-0.2cm}  p_r \, \dot{r} \, ds \right)_{+} \text{exp}  \left (-i  \int^{T}_0 \hspace*{-0.2cm} p_r \, \dot{r} \, ds \right)_{-} \approx 1
\label{in2}
\eea
where the minus sign in front of the second exponent  can be eliminated in favor of the choice $f=-1$. Upon collecting the phase factors on the right hand side, it is easy to see that the exponential part of the flux density across the horizon is indeed given by the integration cycle of the previous section.

\textbf{Quasiperiodic trajectories:} Off the equator and/or for $p_{\theta}(0) \neq 0$, trajectories that cover the horizon area have two periods:  the closure period, $\tilde{T}$, indicating the proper time scale needed for the trajectory to return to its starting point, and a shorter period, $T_1$, representing the time scale for the radial part of the trajectory to make a single round trip across the horizon, without having the whole trajectory to return to its initial point.  For instance for $\theta(0)=\pi/2 \pm \epsilon$ and $p_{\theta}(0) = 0$, the trajectories have a very large closure period, $\tilde{T}$ whereas the first  period, $T_1$ stays close to the period of the purely radial trajectory. In the case where the closure period is finite, the trajectory is said to be multiply periodic or multiperiodic, otherwise rest of the bounded motion is predominantly quasiperiodic, where trajectories keep tracing regions that are infinitesimally close to their starting points, but fail to close on themselves.  

The cycles that yield the tunneling amplitude are chosen in the same way as before but the evaluation of the classical action using quasiperiodic trajectories must be handled with care. This is because the straightforward replacement of the period, $T$ by the closure period, $\tilde{T}$ leads to an undesirable limiting behavior for the tunneling amplitude: the  value of the action, $W_r(\tilde{T})=-i \int_0^{\tilde{T}}p_r \dot{r} \,d s $ grows with increasing $\tilde{T}$,  but $\tilde{T}$ effectively goes to infinity as $\theta$ approaches to $\pi/2$ and so does  $W_r(\tilde{T})$ over the chosen cycle. This is also true for the case where $\theta(0)=\pi/2$ and $p_{\theta}(0)$ smoothly approaches to zero. To remedy this, tunneling treatment of quasiperiodic trajectories should take the first period,  $T_1$ into account.  To incorporate $T_1$ into this picture, we go back to equation  (\ref{effective}) and rewrite the effective action as:
\begin{eqnarray}
\Gamma_{\rm eff}^{\rm scalar}&=& i \,{\rm tr}\, \ln\left[D_\mu^2-m^2\right]=-i\int_0^\infty \frac{d T_1}{T_1} \,{\rm tr}\,e^{-i\, \frac{1}{2} \left( D_{\mu}^2- m^2 \right)\, T_1}
\label{effectiver2}
\end{eqnarray}
where Schwinger parameter is chosen as the first period \cite{wl3}. The exponent in (\ref{effectiver2}) can be recast in a form that includes the closure period such that 
\begin{eqnarray}
\Gamma_{\rm eff}^{\rm scalar}&=& -i\int_0^\infty \frac{d T_1}{T_1} \,{\rm tr}\,e^{-i\, \frac{1}{2} \frac{T_1}{\tilde{T}} \left( D_{\mu}^2- m^2 \right)\, \tilde{T}}
\label{effectiver3}
\end{eqnarray}
Here, it is understood that ordering parameter is chosen as the closure period and the trace is to be performed over the space of classical, bounded trajectories for which the ratio, $ T_1/ \tilde{T}$ is constant. With this in mind the semiclassical approximation for the path integral proceeds in the usual way.  The resulting change in the path summation given by  (\ref{emit}) is that now classical action is weighted by the factor, $ T_1/ \tilde{T}$:
\bea
\sigma \approx  \mathcal{N} 4 \pi M r_+ \int d\omega \, d j  d p_{\theta}\int  \,d \theta \sin{\theta}    \,e^{-i  \frac{T_1}{\tilde{T}}\int_0^{\tilde{T}} p_{r}  \dot{r} d s } 
\label{emit2}
\eea
Here, the angular part of the action, $W_{\theta}=-i T_1/ \tilde{T} \int_0^{\tilde{T}} p_{\theta} \, \dot{\theta} \, ds $ does not contribute to the path summation  because its value remains uniform across both horizons, therefore vanishes on every tunneling cycle. It is worth noting that $W_{\theta}$ on the trajectory can in fact be specified as a function of Carter's constant. Its value remains the same for all the initial values of $\theta(0)$ and $p_{\theta}(0)$, such that the value of $\mathcal{C}$ remains fixed and vanishes as $\mathcal{C}\rightarrow 0$.  

For multiperiodic trajectories, the factor $  \tilde{T} /T_1$ is the winding number (see Fig \ref{f4}), which is defined as the ratio of the radial action evaluated on the cycle to the tunneling exponents given in (\ref{tun}) and (\ref{tuni}).  The appearance of  $ T_1/ \tilde{T}$  ensures that Hawking temperature remains uniform over the horizon area.  The fact that tunneling exponent remains the same over the horizon is well known and can be inferred from the fact that the location of $r_{\pm}$ does not depend on $\theta$, but in the worldline method it follows from a simple reasoning about the limiting behavior of the tunneling amplitude. Here, it is important to note that the averaging of classical action with respect to the periods should not only apply to multiperiodic trajectories, whose closure period is finite, but it must also  be extended to include quasiperiodic trajectories, for which $ \tilde{T} $ effectively goes to infinity. Otherwise, the tunneling interpretation via orbits would only apply to a countable set of points over the horizon area.  This is because the initial conditions for multiperiodic trajectories constitute a set of measure zero in the phase space, or more simply put the majority of the initial conditions gives rise to quasiperiodic orbits. To formally include quasiperiodic orbits in tunneling picture, the exponent must be evaluated in the limit: $\tilde{T} \rightarrow \infty$. In practical terms, this means the radial action must be integrated on the quasiperiodic orbits for large enough $s$, so that the imaginary part of the exponent arising from quasiperiodic oscillations around the initial point of the trajectory averages out to zero. The remaining part of the exponent is purely real and precisely corresponds to the tunneling exponent. In the averaging process, the factor  $\tilde{T}/ T_1$ can be deduced by counting the number of times that $p_r \dot{r}$ grows to its large values in the vicinity of the event horizon.

\begin{figure}
\includegraphics[width=7cm,height=6cm]{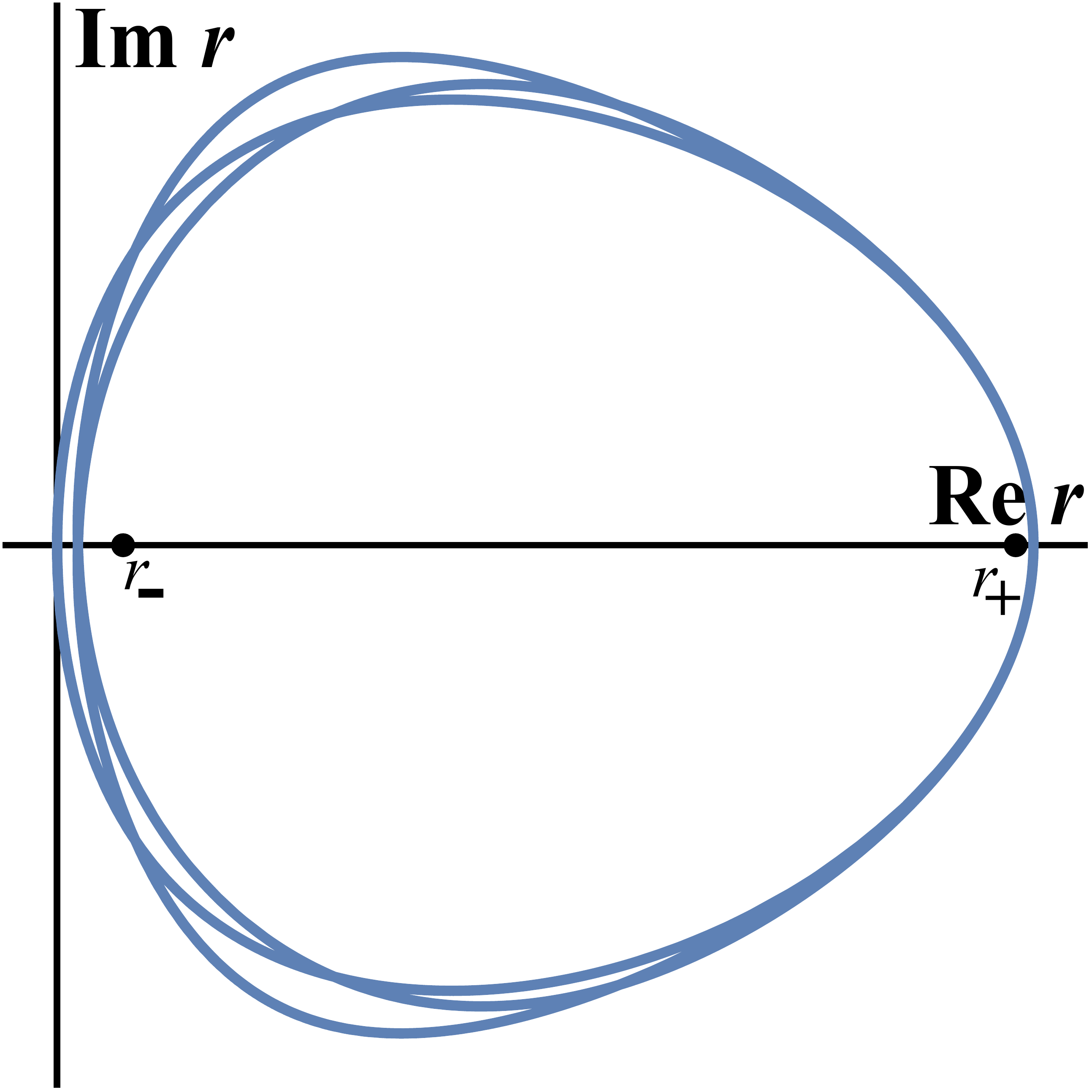}
\includegraphics[width=7cm,height=6cm]{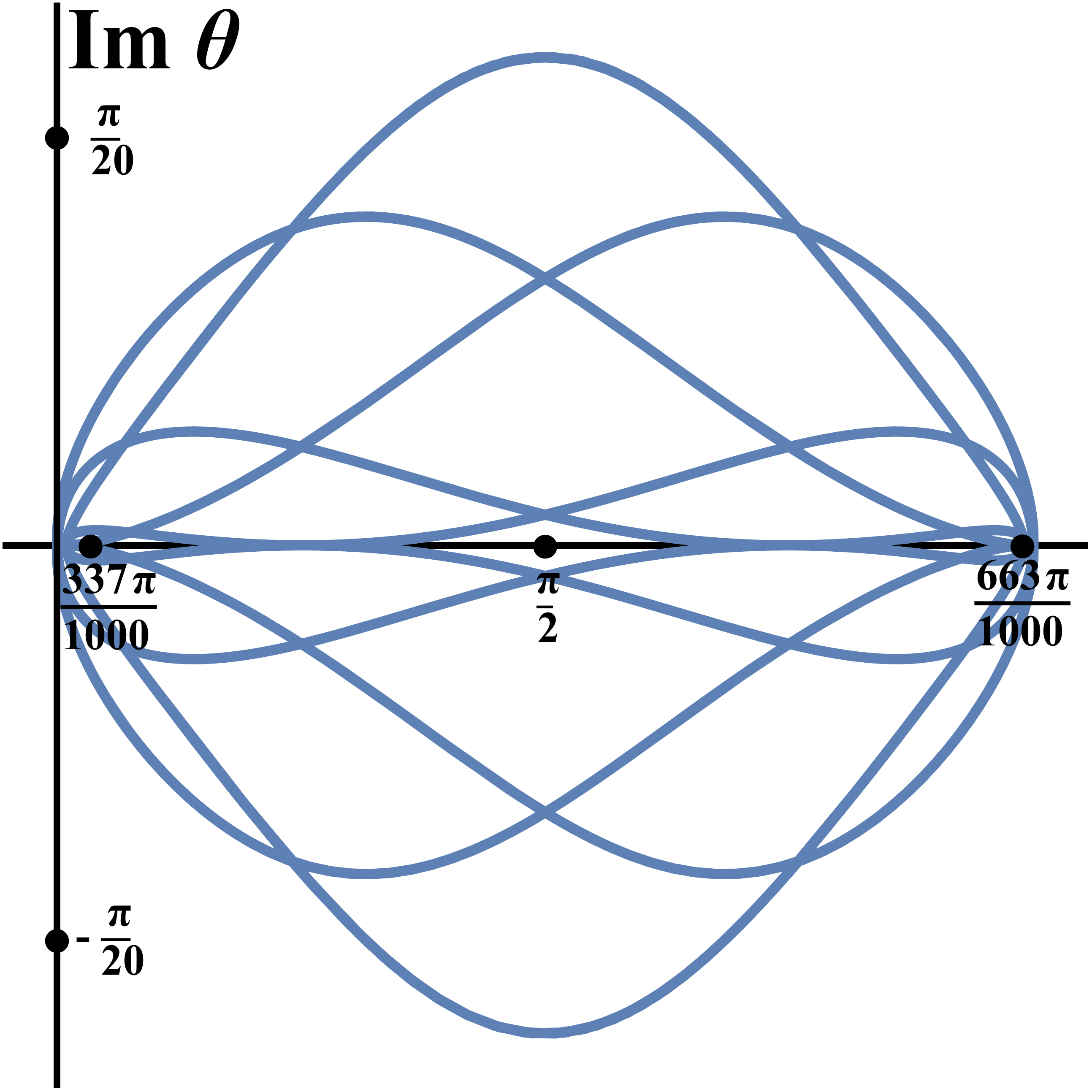}
\caption{A complex multiperiodic trajectory on the complex $r$-plane(left, unscaled) and $\theta$-plane(right) with the winding number, $\tilde{T}/T_1=6$. The radial part of the trajectory makes two traversals until angular part completes the full rotation. The parameters are chosen as: $M=10^5, \, a=9.7 \times 10^4,\, m=1, \, \omega = 10^{-3}, \, j=97, \, \theta(0)=33698 \,\pi/100000 ,  \, p_{\theta}(0)=0 \, \text{and}\, \, d=1$.} 
\label{f5}
\end{figure}

\section{The Geodesic Portrait and Emitted Power} In the resulting picture the horizon area is covered by a dense set of quasiperiodic tunneling orbits. The orbits with same value of $\mathcal{C}$ and conserved momenta  form a circular band of trajectories over the horizon, within the angular interval, $(\theta_i , \, \theta_f)$ and with the corresponding momenta:  $(p_{\theta_i}=0 , \, p_{\theta_f})$. The distribution is symmetric with respect to the equator due to axisymmetry. However such partitioning of the orbits is not unique because the value of $\mathcal{C}$ depends on the chosen value of $d$. Under different  scalings one has the following  relations
\bea
 d'^{\,2} \,\mathcal{C'} \simeq d''^{\,2} \, \mathcal{C}'', \,  \qquad d' p'_{\theta_f} \simeq d'' \, p''_{\theta_f} 
\eea
whereas the angular interval for the partition remains intact to very good approximation:  $\theta'_i \approx \theta''_i, \, \theta'_f \approx \theta''_f$, provided that the inequality
\bea
a^2 m^2/d^2 \gg a^2\omega^2 - j^2/\sin^2{\theta}
\label{ineq}
\eea
holds for the chosen value of $d$. In fact, the angular interval sensitively depends on the ratio, $a/M$. Trajectories within this interval form a equivalence class represented by the scaling relations above. For large momenta saturating (\ref{ineq}),  the angular interval becomes sensitive to the value of momenta. This is because  bounded motion depends sensitively on the location of critical points, which in turn have a sensitive dependence on the momenta and $\theta$. For instance, the critical points, $r^{\pm}_{\text{cr}}$ get closer, merge and recede from one other as complex conjugate pairs as the energy of the tunneling particle increases. Recalling that critical points behave as attractors, the above situation implies that periodic trajectories that lie outside the horizon start migrating inside the horizon with the  increasing energy. In the regime where $\omega > m/d$, there is no periodic trajectory  crossing the real line outside the horizon. A simple way to avoid this difficulty is to scale $m/d$ as a function of momenta such that the location of $r^{+}_{\text{cr}}$ remains fixed at a small but arbitrary distance, $\lambda$ from the horizon. This requires solving: $r^+_{\rm cr}(\mathcal{C} , \, \omega, \, j, \,m/d)=r_+-\lambda $, where $p_{\theta}$ and $\theta$ dependence of $r^+_{\rm cr}$ is encoded by $\mathcal{C}$. Remarkably, this equation can be solved in both BL and Kerr coordinates for the vanishing  $\mathcal{C}$ yielding
\bea
 m^2/d^2 \approx  \,\frac{ 2 M^2  \left(\omega - j \Omega_+ \right)^ 2 } { \lambda \,\sqrt{M^2-a^2}}
\label{scale}
\eea
where $\mathcal{O}({\lambda})$ and higher order terms in the numerator are neglected. Note that  scaling made in (\ref{scale}) also fixes the location of $r^-_{\rm cr}$ to a very good approximation. This means location of poles and critical points  becomes frozen in the momenta, enabling summation over the periodic tunneling cycles up to an arbitrary cut-off.  With this in mind, we may now write the emittance by  using  (\ref{scale}) and the Carter's constant leading to
 \bea
\sigma \approx  \mathcal{N} 4 \pi M r_+ \lim_{\mathcal{C}\rightarrow 0} \int d\omega \, d j  \int  \,d \theta \sin{\theta}  \left[ \mathcal{C}+ \cos^2{\theta} \left( \frac{2 M^2   \left(\omega - j \Omega_+ \right)^ 2}{ \lambda \, \sqrt{M^2-a^2}}a ^2 
+  a^2 \omega^2 - \frac{j^2}{sin^2{\theta}}\right) \right] ^{1/2}  \,e^{- \frac{\left| \omega-\Omega_+ j \right|}{T^{+}_H}  } 
\label{emit3}
\eea
which represents a degraded path summation in the sense that $p_{\theta}$ integration is performed in the limit of vanishing $\mathcal{C}$ only. In this limit the corresponding integration region for $\theta$ now includes the equator with $\theta_i=\alpha, \, \theta_f=\pi-\alpha$. As  $\lambda$ is small enough, (\ref{ineq}) holds therefore the numerical value of $\alpha$ is not sensitive to the value of the momenta. For instance, for $ a/M=0.97$, $\alpha$ saturates the limit: $ \sim 36 \pi /100$ for the decreasing values of $\lambda$. Assuming that integration interval is sufficiently far away from the poles, the first term inside the square root in (\ref{emit3}) dominates, enabling us to safely neglect the remaining terms. Upon performance of $\theta$ integral in the limit $\mathcal{C}\rightarrow 0$, one has
 \bea
\sigma \approx  \mathcal{N}  \frac{ 4 \sqrt{2} \pi M^2 r_+\, a \cos^2{\alpha} }{\sqrt{{ \lambda \,  \sqrt{M^2-a^2}} }} \, \int   \,d\omega \, d j  \, e^ {-\frac {\left|\omega - j \Omega_+ \right|}{T^+_H}}  \, \left|\omega - j \Omega_+ \right|
\label{emit4}
\eea
The factor multiplying the tunneling amplitude has a natural interpretation as the density of states, originating from the Carter's constant.  Consequently, the energy radiated per unit time, per unit area at the horizon can be given  in analogy with the standard black body radiation
\bea
\mathcal{P} & \approx &\mathcal{N} \frac{a M \cos {\alpha} }{ \sqrt{{ 2\lambda \,  \sqrt{M^2-a^2}} }} \, \int   \omega \,d\omega \, d j  \, e^ {-\frac {\left|\omega - j \Omega_+ \right|}{T^+_H}}  \,  \left|\omega - j \Omega_+ \right|
\label{black1}
\eea
where the total area is taken to be the area of the circular band with vanishing $\mathcal{C}$. Above expression is suitable for integration in the deep subradiant and superradiant regime (for a greater accuracy, including the modes $\omega \sim \Omega_+ j$, one should include higher order terms in $\lambda$ in (\ref{scale})).  Now to illustrate the blackbody character of the emitted radiation, we integrate the modes in the subradiant and superradiant regimes up to some cut-off in energy and angular momentum respectively. In subradiant regime, we denote the integration interval for $j$ by $(0, \, \omega \rho )$, where $\rho$ is a constant that must satisfy $\rho\, \Omega_+ \ll 1$ and has the dimension of length. Since there is no canonical choice for $\rho$ we will leave it unspecified. The integration over $j$, followed by the integration of $\omega$ over the region $(0, \, \omega_c)$, where $\omega_c$ designates the unspecified cut-off, ultimately yields the power radiated per unit area as
\bea
\mathcal{P}^{\rm sub} & \approx & \mathcal{N}  \frac{ a M  \cos {\alpha}}{\sqrt{{2 \lambda \sqrt{M^2-a^2}} }} 
 \frac{3\,  \rho\,(2-\rho \,\Omega_+)(T^{+}_H)^{4}}{(-1+
\rho \, \Omega_+)^2} 
\eea
The next order terms neglected in the above expression are all suppressed exponentially $\sim \mathcal{O}(e^{-\omega_c/T^+_{H}})$. The integration of superradiant modes is carried out in a similar manner except that the order of integration is reversed. In specifying the upper limit for $\omega$, we use the same parameter $\rho$  as before. Since $\rho$ has the dimension of length and $j$ is unitless we identify $j \rho\,\Omega^2_+$  as the end point of $\omega$ integration.  Integrating over momenta with the cut-off, $j_c$ yields
\bea
\mathcal{P}^{\rm sup} & \approx & \mathcal{N}  \frac{ a M \cos {\alpha}}{ \sqrt{{2 \lambda \sqrt{M^2-a^2}} }} 
 \frac{3  \, \rho^2 \Omega_+ (T^{+}_H)^{4}}{(-1+
\rho \, \Omega_+)^2} 
\eea
where the neglected terms are of the order  $\sim \mathcal{O}(e^{-j_c \Omega_+/T^+_H})$. Adding the two terms together the total flux reads
\bea
\mathcal{P}^{\rm tot} & \approx & \mathcal{N}  \frac{ a M \cos {\alpha}}{ \sqrt{{2 \lambda \sqrt{M^2-a^2}} }} 
 \frac{6   \rho \, (T^{+}_H)^{4}}{(-1+
\rho \, \Omega_+)^2} 
\label{tsig}
\eea
The path summation for $\mathcal{C} \neq 0$ is more involved and we will not attempt to obtain a closed form expression  here but the approach is identical to $\mathcal{C}=0$ case. The difference is that the magnitude of $\lambda$ now fixes the scale of $\mathcal{C}$ and $p_{\theta}$ through
\bea
p_{\theta}= \sqrt{\mathcal{C}+\cos^2{\theta} \left( \frac{a^2 m^2}{d^2(\omega, \, j, \, \mathcal{C}, \lambda) }+ a^2\omega^2 -\frac{j^2}{\sin^2{\theta}}\right)}
\label{azimom}
\eea 
As before, $d^2(\omega, \, j, \, \mathcal{C}, \lambda)$ is the solution of $r^+_{\rm cr}(\mathcal{C} , \, \omega, \, j, \, d)=r_+-\lambda $. In order to specify the band, $(\theta_{i}, \theta_{f})$, one first picks an initial point $\theta_{i}$ for the chosen value of  the conserved momenta and $\lambda$. The corresponding value of $\mathcal{C}$ is given by (\ref{azimom}), where $p_{\theta_{i}}$ is required to vanish. The final point $\theta_{f}$, beyond which the trajectories for the given  value of $\mathcal{C}$ becomes unbound  can be obtained by direct integration of the orbits. The orbits covering the horizon can be organized into bands labeled by $\mathcal{C}$ by repeating this procedure for different values of $\theta_{i}$. The emergent qualitative property of such partitioning is that the thickness of the bands decrease as $\theta_i$ gets closer to the poles. At the poles,  worldline prescription fails to give bounded trajectories for the non-vanishing angular momentum because Carter's constant in spheroidal coordinates diverges.  The end result of the path summation  for a chosen partition can be shown to depend only on the initial and final angles, not on the specific value that $\mathcal{C}$ takes, for the fact that $\mathcal{C}$ can algebraically be related to the $\theta_{i}$ for the vanishing $p_{\theta_{i}}$.

\section{Discussion}

The main result behind (\ref{tsig}) is that  the spectrum is predominantly in thermal character $\sim (T^{+}_H)^{4}$ as long as the phase portrait of the system remains uniform. This means   $\lambda$ appearing in  (\ref{tsig}) is much smaller than the horizon length scale $\sim M$ and its value approximately stays the same with respect to the momenta. Here, $\lambda$  can intuitively be viewed as the tunneling/penetration depth  because it gauges the distance scale over which the radial action makes a phase jump. At energies above some cut-off scale, the momentum dependence of $\lambda$ prompts nonlinearity  in the density of states, thus signaling deviations from the thermal spectrum. The fact that deviations should occur may be seen in an alternative way. Note that period of the radial orbit for the vanishing momenta is simply $T=2 \pi M d /m$. Because of the scaling made in (\ref{scale}), the period of the orbits shrinks down to arbitrarily small scales with the increasing energy of the tunneling flux.  This is reminiscent of the transplanckian problem for the energetic modes, where the usual semiclassics breaks down and the backreaction on the metric must be taken into account. In addition to these observations,  the fact that the prefactor depends on the black hole parameters shows that emitting body deviates from a perfect blackbody, which is an idealization.  This observation remains valid even if we ignore the factor $M/\sqrt{\lambda \sqrt{M^2-a^2}}$, which was brought by the scaling argument that we made; the remaining factor of  $a$ in the prefactor shows the influence of background geometry on the density of states.

To conclude, we have shown that tunneling treatment of massive particles from the Kerr event horizon can be given by using complex geodesics. In fact, the use of such complex trajectories becomes essential, if one wants to capture the tunneling mechanism throughout the whole horizon area via the orbit picture. In this process, the averaging of the action with respect to the periods emerges as the necessary ingredient enabling the tunneling exponent to have the right limit when $\theta \rightarrow \pi/2$ and $p_{\theta} \rightarrow 0$. As a direct result of this manipulation, the value of $ -i\, T_1 / \tilde{T} \,W_r$  on the cycle does not  depend on the winding number. This property indicates $T^{\pm}_H$ is invariant under modular transformations that connect cycles with differing winding numbers over the horizon or simply put, the event horizon is a \textit{uniform tunneling surface}.  An important aspect of worldline formalism is that integration cycles automatically give the tunneling probability for the inner horizon as well as the outer horizon. But it is not obvious that one should worry about the consequences of  inner horizon tunneling. If the semiclassical picture still remains valid there, one may argue that the size of the inner horizon should undergo fluctuations as a result of  steady  emission and reabsorption. In the time dependent picture the rate of the fluctuations evidently depends on $T^{-}_H$, whereas the size of $r_-$ shrinks at a rate determined by $T^{+}_H$.

\bigskip

I would like to thank Gerald Dunne and Mithat \"{U}nsal for the comments on the manuscript, and  Bayram Tekin for pointing out the reference\cite{hagels}. I would also like to thank Turan Birol for discussions and the anonymous referees for the  comments.

\bigskip

\appendix
\section{Schwinger effect}

Here we give the details for the integration of the zero modes. We begin by considering the tunneling exponent. Recalling (\ref{actions}) and (\ref{clt}), the proper time integration of the  total exponent in (\ref{scalar1}) yields:
\bea
\Delta = -m^2 s /2 - \frac{m^2 (2+\gamma^2)\, s }{2 \gamma^2}+ \frac{2 m}{2 \gamma k}  \,\text{ArcTan}\left[
\frac{\text{Tan}\left[\frac{m s \sqrt{1+\gamma^2}k}{\gamma} \right]} {\sqrt{1+\gamma^2}}\right]
\eea
By taking into account the branch points at $s=\frac{(\pi/2 +2\pi n) \gamma}{m k \sqrt{1 +\gamma^2}}, \, n \in \mathbb{Z} $, we may take the limit, $s\rightarrow T_c$ giving
\bea
\Delta = -m^2 T_c /2 + \frac{2 \pi m}{\gamma k} - \frac{m \pi (2+ \gamma^2)}{\gamma k \sqrt{1+\gamma^2}}
\label{action2}
\eea
Note that here $m^2$ and $T_c$ are conjugate variables. This mean mass dependence of the second and third terms above must be eliminated by using the definitions of $\gamma$  and $T_c$  so that action, $S[x^\mu(s); T]$ is solely given in terms of $T_c$. Doing so we have: 
\bea
\Delta = -m^2 T_c/2 -\frac{\left( E_0 q \,T_c-2\pi \right)^2}{2 T_c k^2}
\eea
which shows that on the saddle point $\partial \Delta / \partial{T_c}$ vanishes identically. The contribution coming from the second variation can readily be calculated:
\bea
 \sqrt{\frac{2\pi}{ 
 \left| \partial^2 \Delta / \partial T_c^2 \right| }} =\frac{ 2 \pi }{m\, \sqrt{m k}} \frac{\gamma^{3/2}}{\left(1+\gamma^2\right)^{3/4}}
\label{sadt}
\eea 
The van-Vleck determinant in (\ref{pre}) is taken into account by considering the second order fluctuations to the action: $S[x^\mu(s); T_c]=S[x^\mu_{\text{cl}}(s); T_c]+ \delta^2 S[x^\mu_{\text{cl}}(s); T_c]$, which is expanded around the classical trajectory such that: $x^{\mu}(s)=x^{\mu}_{\text{cl}}(s)+\eta^{\mu}(s)$. The secondary action, $ \delta^2 S[x^\mu_{\text{cl}}(s); T_c]$ can readily be given as:
\bea
\delta^2 S[x^\mu_{\text{cl}}(s); T_c]=\int \eta^{\mu}(s) \Lambda_{\mu\nu} \eta^{\nu}(s) ds
\eea
where the fluctuation operator reads :
\bea
\Lambda_{\mu\nu}\equiv  -\frac{d}{ds}\left(P_{\mu\nu} \frac{d}{ds} + Q_{\nu\mu}\right) + \left( Q_{\mu\nu}\frac{d}{ds} + R_{\mu\nu} \right)
\eea
with,
\bea
P_{\mu\nu}=\frac{\partial^2 L}{\partial \dot{x}_{\mu}\partial \dot{x}_{\nu}}, \quad Q_{\mu\nu}=\frac{\partial^2 L}{\partial x_{\mu}\partial \dot{x}_{\nu}}, \quad R_{\mu\nu}=\frac{\partial^2 L}{\partial x_{\mu}\partial x_{\nu}}
\eea 
The fluctuation fields  $\eta$ satisfy Morse's boundary problem, also known as the Jacobi equation \cite{morse}:
\bea
 \Lambda_{\mu\nu} \eta_{\nu}=\lambda \eta_{\mu}, \quad \eta_{\mu}(0)=\eta_{\mu}(T_c)=0
\label{morse1}
\eea
Using a suitable orthonormal basis expansion, $\eta=\sum_n a_n u_n$, one may perform the Fresnel integrals over $a_n$. Taking into account the metric signature, the product of eigenvalues resulting from  integration in $x_0-x_1$ plane is normalized according to \cite{levit1}:
\bea
 -\frac{i}{2 \pi T_c} \left | \frac{\prod_{n} \lambda^{\text{free}}_n}{\prod_{n} \lambda_n} \right| = -\frac{i}{2 \pi T_c} \left | \prod_{n}  \frac{\lambda^{\text{free}}_n} {\lambda_n} \right |
\label{norm}
\eea
The appearance of $\lambda^{\text{free}}_n$ above can be viewed as the normalization of the measure $d a_n$. The overall normalization factor ensures that the result of integration coincides with the free particle prefactor in the vanishing field limit.  With Lagragian given in (\ref{actions}), we  may write the fluctuation operator explicitly ($\mu , \, \nu=0,1,2,3 $):
\bea
\Lambda_{\mu\nu} = \begin{pmatrix}
  -i\frac{d^2}{d s^2}-q\, \dot{x}^1 \partial^2_0 A_1 & -q \, \partial_0 A_1 \frac{d}{d s} &0 &0 \\\\
   q \, \dot{x}^0 \partial^2_0 A_1  + q \, \partial_0 A_1 \frac{d}{ds}&  i \frac{d^2 }{d s^2} &0 &0\\\\
0& 0 &  i \frac{d^2 }{d s^2}  & 0 \\\\
0& 0 &  0 & i \frac{d^2 }{d s^2} 
\end{pmatrix}
\label{fluc1}
\eea
Note that above matrix is in block diagonal form. The path integral can be  factorized into two parts, where the determinant of the lower block is just the free particle prefactor, encoding second order fluctuations perpendicular to $x^0 - x^1$ plane. Then we may write the total determinant as:
\bea
 \left (\sqrt{\text{det}\, \frac{\partial p_{\mu}(x(0))}{\partial x^{\nu}(T) }} \right)_{T=T_c}  =  -\frac{i}{\left( 2 \pi T_c \right)^2}\left| \prod_{n}  \frac{\lambda^{\text{free}}_n} {\lambda_n}\right|
\eea
where now product term represents the determinant of the reduced fluctuation operator, $\tilde{\Lambda}_{\mu \nu}, \, \mu,\nu=0,1$. In order to proceed, we will  make use of an important theorem from\cite{levit2}, where in the continuum limit  the ratio of  the eigenvalues is given by:
\bea
\left| \prod_{n} \frac{\lambda^{\text{free}}_n} {\lambda_n}\right| \equiv \left| \frac{\text{det}\, \left[\tilde{\eta}^{(\nu)\, \text{free}}_{\mu}(s) \right]} {\text{det} \,\left[ \tilde{\eta}^{(\nu)}_{\mu}(s) \right]}\right|_{s=T_c}^{1/2}
\label{van}
\eea
such that $\tilde{\eta}(s) $ is the solution of the corresponding initial value problem:
\bea
\tilde{\Lambda}_{\mu \nu} \tilde{\eta}_{\nu}(s)=0, \quad \tilde{\eta}^{\nu}_{\mu}(0)=0, \quad \dot{\tilde{\eta}}^{\nu}_{\mu}(0)=g_{\mu\nu}
\label{jbc}
\eea 
The zero modes that solve the free particle  limit of the Jacobi equations can be readily found, yielding a determinant factor of $i T_c$.  Remarkably, for the nontrivial part,  the algebraic solutions satisfying the above initial conditions can also be found by making use of  the classical equations of motions, ultimately yielding:
\bea
\tilde{\eta}^{0} = i m \begin{pmatrix}
\dot{x}_{\text{cl}}^0(s) I_1 \\\\
  \dot{x}_{\text{cl}}^1(s) I_1- I_2& 
\end{pmatrix}, \quad 
\tilde{\eta}^{1} =-\begin{pmatrix}
\dot{x}_{\text{cl}}^0(s) I_2 \\\\
  \dot{x}_{\text{cl}}^1(s) I_2- m^2 I_1& 
\end{pmatrix}  
\label{jf1}
\eea
where, 
\bea
I_1 = \int_0^{s} \frac{1}{\left(\dot{x}_{\text{cl}}^0 (s')\right)^2} d s', \quad I_2 = \int_0^{s} \frac{\dot{x}_{\text{cl}}^1(s')}{\left(\dot{x}^0_{\text{cl}}(s')\right)^2} ds'
\label{jf2}
\eea
Note that the determinant factor $({\text{det} \,[ \tilde{\eta}^{(\nu)}_{\mu}(s)]})^{-1/2}$ in (\ref{van}) vanishes at the conjugate points, where $\dot{x}_{\text{cl}}^0=0$. These points are located at $x_{\text{cl}}^0(T_c /4)$ and $x_{\text{cl}}^0(3T_c /4)$, coinciding with the critical points of the action.  The important thing to note here is that because the determinant was given inside the absolute value from the beginning, the sign change of the radicand  must be accounted by a overall factor of $e^{-i \pi/2}$ at each conjugate point. The number of times radicand changes the sign along the trajectory gives the Morse index of trajectory. Here, the Morse index is simply 2.   Now returning back to (\ref{van})  the Van-Vleck determinant can be given with the aid of (\ref{jf1}) and (\ref{jf2}):
\bea
 \left (\sqrt{\text{det}\, \frac{\partial p_{\mu}(x(0))}{\partial x^{\nu}(T) }} \right)_{T=T_c}=-i m^2 k \frac{(1+\gamma^2)^2}{16 \pi^4 \gamma^4}
\label{van2}
\eea
The final piece of (\ref{scalar1}) to look at is the volume integral over the initial points of the  trajectory. Because  the equations of motion are invariant under proper time translations,  each point located on the trajectory is a legitimate starting point. The important consequence of this fact  is  that  volume integral contains a multiplicity factor given by the path length.  This becomes evident upon making the substitution, $d\, x^{0}(0)\rightarrow \dot{x}^{0}(0) ds $  and integrating over $s$, yielding $i m \, T_c$. The factor $T_c$ here cancels against the factor  $1/T_c$ appearing in (\ref{pre}), leaving an overall normalization factor of $i m$ for the zero modes. Taking this normalization factor and the Morse index into account and collecting all the terms in (\ref{action2}), (\ref{sadt}) and (\ref{van2})  together, the imaginary part of  the one-loop effective action per unit volume leads to the result given by (\ref{eff1}).


\begin{thebibliography}{999}
\bibitem{dewittc}
In performing the discretization we use the resolution of identity as $\int 
 \ket  x \bra x d^4x $ so that transition amplitude between initial and final points on the trajectory is given as a scalar density. Accordingly, the basis vector $ \ket  x$ transforms as vector density of weight $\frac{1}{2}$. Alternatively, using: $\int \ket x \bra x \sqrt{-g} \, d^4 x $ with the normalization $\braket {x}{p}= \text{exp} \left(-i p_{\mu} x^{\mu}\right)/(2\pi (-g)^{1/4})$ results in the cancellation of the $\sqrt{-g}$ factor in the volume element. The net effect of using the the latter choice is that the overall prefactor multiplying the momentum integrals in (\ref{black1}) must be replaced by : $ \mathcal{P}\, a(1-\sin{\alpha})/ \left(2
\sqrt{2}\pi \cos{\alpha} \,  r_+ \sqrt{\lambda \sqrt{M^2-a^2}} \right )$
\end{thebibliography}
\end{document}